\documentclass[letterpaper,oneside,12pt]{article}
\usepackage[dvipsnames]{xcolor}
\usepackage{mathrsfs}
\usepackage{appendix}
\usepackage{natbib}
\usepackage{extarrows}
\usepackage[figuresright]{rotating}
\usepackage{graphicx,setspace,lscape,longtable,epstopdf,xr}
\usepackage{amsmath,amsthm,amssymb,color}
\allowdisplaybreaks[4]
\usepackage{pdfpages}
\usepackage{float}
\usepackage{bm}
\usepackage{chngcntr}
\usepackage{geometry}
\usepackage{authblk}
\usepackage[normalem]{ulem}

\usepackage{etoolbox}
\AtBeginEnvironment{align}{\vspace{-2\abovedisplayskip}}
\AtEndEnvironment{align}{\vspace{-2\belowdisplayskip}}
\AtBeginEnvironment{align*}{\vspace{-2\abovedisplayskip}}
\AtEndEnvironment{align*}{\vspace{-2\belowdisplayskip}}
\AtBeginEnvironment{equation}{\vspace{-1.5\abovedisplayskip}}
\AtEndEnvironment{equation}{\vspace{-1.5\belowdisplayskip}}
\newcommand{\compressmath}{%
  \setlength{\abovedisplayskip}{3pt}
  \setlength{\belowdisplayskip}{3pt}
  \setlength{\abovedisplayshortskip}{3pt}
  \setlength{\belowdisplayshortskip}{3pt}
}
\usepackage{titlesec}

\titlespacing*{\section}{0pt}{4pt}{2pt}
\titlespacing*{\subsection}{0pt}{3pt}{1pt}
\titlespacing*{\subsubsection}{0pt}{2pt}{1pt}

\newtheorem{lemmasection}{Lemma}[section]
\def\bels{\begin{lemmasection}}
\def\eels{\end{lemmasection}}	

\usepackage{multirow}
\usepackage{ragged2e}
\usepackage{todonotes}
\usepackage{hyperref}
\hypersetup{
colorlinks=true,
linkcolor=blue,
citecolor=blue,
urlcolor=blue}
\usepackage{booktabs}
\usepackage{pdflscape}
\usepackage{algorithm}
\usepackage{algpseudocode}
\usepackage{subfigure}
\usepackage{enumerate}
\algrenewtext{Repeat}{\textbf{Repeat: }}
\algrenewtext{Until}{\textbf{Until }}
\bibpunct{(}{)}{;}{a}{,}{,}

\def\spacingset#1{\renewcommand{\baselinestretch}%
{#1}\small\normalsize} \spacingset{1}

\newtheorem{theorem}{Theorem}
\newtheorem{lemma}{Lemma}
\newtheorem{proposition}{Proposition}

\def\p{\prime}

\def\ve{\varepsilon}

\def\beq{\begin{equation}}
\def\eeq{\end{equation}}
\def\beqr{\begin{eqnarray}}
\def\eeqr{\end{eqnarray}}
\def\beqrs{\begin{eqnarray*}}
\def\eeqrs{\end{eqnarray*}}
\def\bet{\begin{theorem}}
\def\eet{\end{theorem}}
\def\bel{\begin{lemma}}
\def\eel{\end{lemma}}
\def\bep{\begin{proposition}}
\def\eep{\end{proposition}}
\def\bg{\begin{figure}[tbph]\begin{center}}
\def\eg{\end{center}\end{figure}}

\def\bc{\begin{center}}
\def\ec{\end{center}}

\def\IC{\mbox{IC}}

\def\cB{\mathcal{B}}

\newtheorem{remark}{Remark}
\newtheorem{corollary}{Corollary}

\def\wt{\widetilde}

\def\wh{\widehat}

\def\M{\mathbf{M}}
\def\W{\mathbf{W}}

\def\ol{\overline}

\def\mC{\mathcal C}

\def\mD{\mathbb D}
\def\mQ{\mathcal Q}

\def\cG{\mathcal G}
\def\mR{\mathbb{R}}

\def\mL{\mathcal L}

\def\mS{\boldsymbol{S}}

\def\mT{\mathcal T}
\def\bH{\mathbb H}

\def\mZ{\mathbb{Z}}
\def\bx{\mathbf{x}}

\def\Z{\mathbf{Z}}

\def\bZ{\mathbf{Z}}
\def\bV{\mathbf{V}}

\def\bv{\mathbf{v}}

\DeclareMathOperator*{\argmin}{arg\,min}

\newcommand{\balpha}{\boldsymbol{\alpha}}
\newcommand{\bgamma}{\boldsymbol{\gamma}}
\newcommand{\bbeta}{\boldsymbol{\beta}}

\def\bDelta{\boldsymbol{\Delta}}

\newcommand{\RNum}[1]{\uppercase\expandafter{\romannumeral #1\relax}}

\def\bZ{\mbox{\boldmath $Z$}}
\def\Z{\mbox{\boldmath $Z$}}

\def\bz{\mbox{\boldmath $z$}}
\def\z{\mbox{\boldmath $z$}}

\def\bH{\mathbf{H}}

\def\W{\mathbf{W}}

\def\bW{\mathbf{W}}

\def\mD{\mathcal{D}}

\def\bg{\mbox{\boldmath $g$}}

\def\bs{\mbox{\boldmath $s$}}
\def\s{\mbox{\boldmath $s$}}

\def\p{\textup{p}}

\def\alg{\textup{alg}}

\def\defeq{\stackrel{\mathrm{def}}{=}}  % for definitions

\def\boxit#1{\vbox{\hrule\hbox{\vrule\kern6pt\vbox{\kern6pt#1\kern6pt}\kern6pt\vrule}\hrule}}

\numberwithin{equation}{section}

\newcommand{\blind}{1}
\begin{document}
\def\spacingset#1{\renewcommand{\baselinestretch}%
{#1}\small\normalsize} \spacingset{1}

\if1\blind
{
  \title{\bf Consistent Selection of the Number of Groups in Panel Models via Cross-Validation}
    \author[1]{Zhe Li}
    \author[2]{Xuening Zhu}
    \author[3]{Changliang Zou\thanks{Changliang Zou and Xuening Zhu are the     corresponding authors.}}

    \affil[1]{School of Data Science, Fudan University, China}
    \affil[2]{School of Management, Fudan University, China}
    \affil[3]{School of Statistics and Data Science, Nankai University, China}
  \date{}
  \maketitle
} \fi

\if0\blind
{
  \vspace*{2cm}
  \begin{center}
    {\LARGE\bf Consistent Selection of the Number of Groups in Panel Models via Cross-Validation}
\end{center}
\vspace*{3.5cm}
} \fi

\vspace{-1cm}
\begin{abstract}
Group number selection is a key problem for group panel data modeling.
In this work, we develop a cross-validation (CV) method to tackle this problem.
Specifically, we split the panel data into two data folds on the time span with a buffer zone, with  group structure preserved for individuals.
We first estimate the group memberships and parameters on one data fold,
then plug in the estimates and utilize the other data fold to evaluate a designed criterion.
Subsequently, the group number is estimated by minimizing the average criterion across all data folds.
 The proposed CV method has two advantages compared to existing approaches.
 First, the method is totally data-driven; thus no further model-specific tuning parameters are involved.
 Second, the method can be flexibly applied to a wide range of panel data models.
 Theoretically, we establish the estimation consistency by taking advantage of the optimization process on the training data fold.
 Experiments are carried out with a variety of synthetic datasets and panel models to further illustrate the advantages of the proposed method.
 Lastly, the CV method is employed to analyze the heterogeneous patterns of stock volatilities in the Chinese stock market during the 2008 financial crisis.
\end{abstract}

\noindent%
\spacingset{1.9}

\section{Introduction}

Panel data modeling is an important research area in statistics and econometrics \citep{arellano2003panel, ke2015homogeneity,ke2016structure,ando2017clustering, hsiao2022analysis}, as it captures information from both temporal and cross-sectional dimensions.
A critical problem in panel data modeling is to characterize the individual heterogeneity resulting from distinct backgrounds and individual characteristics \citep{bai2014theory,ke2015homogeneity,li2016panel,fan2018heterogeneity,hong2023profile}.
In this regard, latent group structures have recently received considerable attention~\citep{ke2015homogeneity,su2019sieve,gu2019panel,zhu2023simultaneous,mehrabani2023estimation}.
The key assumption in group panel data modeling is that the individuals
within the same group share the same set of model coefficients \citep{ke2015homogeneity,su2016identifying,fang2023group}.
With this specification, research has shown that the latent group structure can capture flexible unobserved heterogeneity of panel data while retaining a parsimonious model form and desirable statistical efficiency.

While group panel data models have proved to be useful, their practical performance is significantly contingent on the specification of the group number.
If the group number is under-specified, one will end up with an underfitted model
with inferior performance.
On the other hand, over-specification of the group number may result in an overfitted model with suboptimal estimation efficiency.
Consequently, a critical problem for the group panel data models is how to determine the number of groups.

For selecting the number of groups/clusters, two common approaches are widely used.
The first involves the utilization of an information criterion (IC) that integrates both model fitness and model complexity.
Methods based on IC are extensively utilized in the literature for tuning parameter selection in model selection tasks \citep{schwarz1978estimating,hurvich1998smoothing,wang2009shrinkage,zhang2010regularization}.
For instance, \cite{naik2007extending} extended the Akaike information criterion to mixture regression models for selecting the number of mixture components.
\cite{li2016panel} employed a specific information criterion to determine the number of structural breaks.
\cite{hu2020corrected} proposed a corrected BIC to determine the number of communities in the community detection task.
In the analysis of grouped panel models, \cite{lin2012estimation} proposed a modified BIC for linear panel models.
\cite{bonhomme2015grouped} used a BIC based criterion for grouped fixed-effects models.
\cite{liu2020identification} designed a PC criterion for panel models with individual fixed effects.
Theoretically, the IC based methods can consistently select the true group number under certain conditions.
However, to implement the IC based methods, one still needs to specify certain model-specific tuning parameters, whose choice may depend on the specific model and error distribution, making the procedure subjective and potentially unstable.
The second one is the hypothesis testing-based method \citep{tibshirani2001estimating,onatski2009testing,choi2017selecting}.
This type of method formulates the group number estimation problem into a sequential hypothesis testing procedure \citep{hardy1996number,lo2001testing,bickel2016hypothesis}.
In the investigation of panel models with group structures,
\cite{lin2012estimation} used a $t$ test to test the homogeneity of the data.
\cite{lu2017determining} proposed a residual-based Lagrange multiplier-type test to determine the group number for linear group panel data models.
However, the hypothesis testing-based methods are usually restricted to the linear panel models
and cannot provide a unified group number estimation framework with general model forms.

\begin{figure}[htbp]
  \centering
  \includegraphics[scale=0.4]{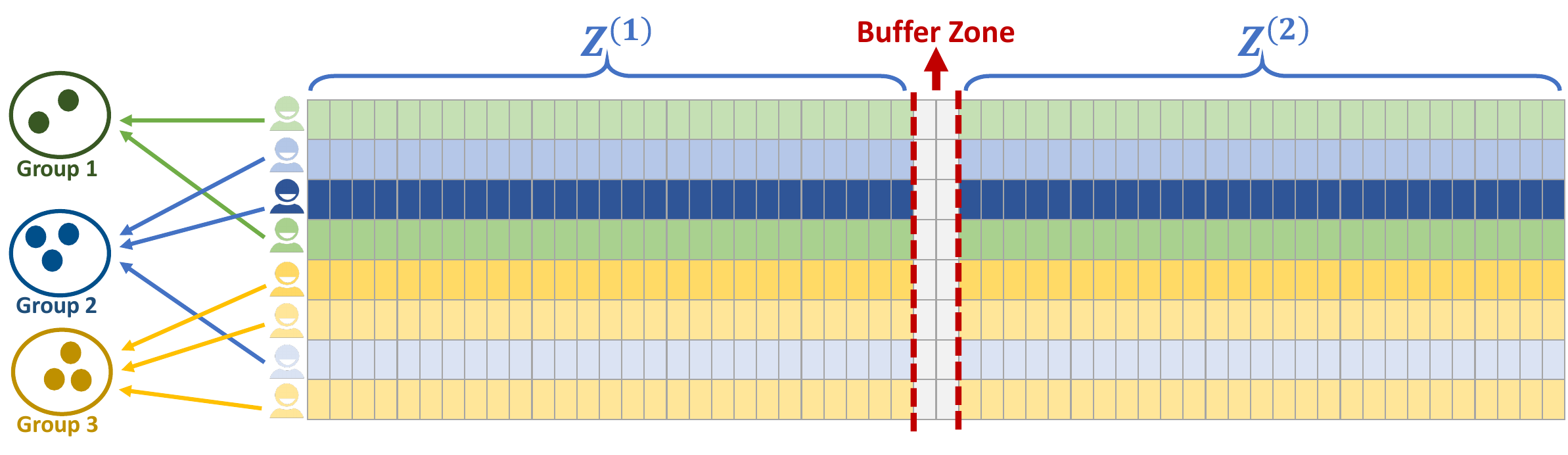}
  \caption{\small\it Schematic diagram of the panel data splitting mechanism.}\label{fig:split}
\end{figure}

In this work, we propose a unified group number estimation method via cross-validation for general panel data models.
The method allows for entirely data-driven implementation, eliminating the need for specifying model-specific tuning parameters.
Specifically,
we first split the panel data on the time span into two folds, and a time gap is specified between the two data folds as a buffer zone to accommodate temporal dependence.
We use the first data fold for model training and the second for model testing, and then swap their roles.
This splitting method  maintains the group structure of individuals of the panel data; see Figure \ref{fig:split} for illustration.
Subsequently, we estimate the group structure and model parameters based on the training data, given a specified group number $G$.
Then, we evaluate a designed criterion with the testing data by approximating the loss with a local quadratic function.
Finally, the determination of the group number is achieved by minimizing the average criterion over two data folds.

The idea is in spirit similar to the $K$-fold cross-validation (CV) method, which is widely used as a common practice to select tuning parameters (or candidate models) or
evaluate the prediction performance of {a wide range of machine learning methods} \citep{wang2007unified,hastie2009elements,wang2007tuning}. %,?
{For model selection, the CV method is typically used to select important tuning parameters for machine learning models, which can yield the best prediction accuracy. 
For example, it can be applied to select the penalty level for the regularized regression models \citep{tibshirani1996regression,wang2007regression}.
In model assessment, the CV method is used to quantify the prediction accuracy \citep{bates2024cross} or make valid statistical inferences \citep{fan2004crossvalidation}.
For example, \cite{fan2012variance} proved that a consistent variance estimation can be obtained for ultrahigh dimensional regression model with the CV technique.
For the group panel data model, the group number $G$ can be treated as a tuning parameter, therefore our task is to utilize the CV method for selecting $G$ instead of assessing model predictions.

}

However, existing theoretical properties have shown that the $K$-fold CV method tends to select an overfitting model when applied to model selection tasks \citep{shao1993linear,wang2007tuning}.
{Nevertheless, we find that it is not true with our task.
Notably, the estimation consistency of group number $G$ can be established under our scenario. 
This is mainly because we consider a two-dimensional panel data setting and we conduct the data splitting on the time span.
As shown by Figure \ref{fig:split}, the splitting on the time span allows us to preserve the group structure on all data folds, and this is the key to establishing the selection consistency by connecting to the optimization procedure with the training data fold.}
%   \sout{{\red We also provide a proof roadmap to illustrate our basic idea.}}
A similar procedure has been adopted by \cite{zou2020consistent} to select the number of change points in the change point detection area.
We also refer to \cite{wang2010consistent,chen2018network,lei2020cross,rabinowicz2022cross} for the relevant literature using CV methods for model selection and relevant inference tasks.
However, to the best of our knowledge, the CV method has not been introduced to group number estimation in the group panel models.

\label{l4p:method_general}
In particular, our method can be applied to a wide range of panel models, including linear panel models \citep{lin2012estimation} and nonlinear panel models, such as probit panel model \citep{su2016identifying}, logit panel model \citep{liu2020identification},
as long as the corresponding loss function can be approximated
with a local quadratic function form \citep{wang2007unified,zou2020consistent,zhu2021least}.
Consequently, it provides a flexible and unified parametric solution for selecting the group number in group panel data models.
Our theoretical framework establishes the connection between the  estimation consistency and the optimization procedures \citep{lin2012estimation,bonhomme2015grouped,liu2020identification}.
We further evaluate the validity of the CV method when applied to a linear panel model with nonstationary covariates \citep{phillips1999linear,tu2017forecasting}, and a classical interactive effects model \citep{bai2009panel,miao2020panel} in Appendix \ref{sec:model_interactive}.
Numerical evidence suggests that the CV method can still yield satisfactory performance  with minor adaptations, illustrating a considerable potential for extensions of the proposed CV method.

The rest of the article is organized as follows.
In Section \ref{sec:selection_with_CV}, we provide  a selection criterion for group number estimation, offering a detailed exposition of the procedure employed for the estimation.
Theoretical properties on asymptotic selection consistency
are established in Section \ref{sec:theoretical_properties}.
Subsequently, Section \ref{sec:model_fixed} extends the proposed selection criterion to group panel models with fixed effects.
Numerical studies are presented in Section \ref{sec:numerical_study} and
an empirical study of stock volatilities
in Chinese stock market during the 2008
financial crisis
 is presented in Section \ref{empirical_study_main}.
Finally, we conclude the article with a discussion in Section \ref{sec:conclusion}.
Additional numerical studies, model extensions, proofs,  and technical lemmas can be found in the Appendices \ref{appendixA}--\ref{sec:numerical_study_appendix}.

\enlargethispage{2\baselineskip}
\section{Selection with Cross-Validation}\label{sec:selection_with_CV}
\subsection{Model and Notation}\label{subsec:model_notation}
\pagebreak[4]

Let $Y_{it}\in\mathbb{R}$ be the response variable and $\bx_{it} \in \mathbb{R}^p$ be
the associated $p$-dimensional covariate vector collected from the  $i$th ($1\le i\le N$) individual at the $t$th ($1\le t\le T$) time point.
The panel data are denoted as $\bZ = \{\bz_{it}: i=1,\ldots,N; t=1,\ldots,T\}$ with $\bz_{it} = (Y_{it}, \bx_{it}^\top)^\top$. 
Suppose that the $N$ individuals are divided into $G_0$ groups, where individuals in the $g$th group ($1\le g\le G_0$) share the same regression coefficient vector $\bbeta_g^0 \in \mathbb{R}^p$. 
For each individual $i$,
denote ${g_i^0} \in \{1,2,\ldots,G_0\}$ as its true group membership. %thus $\bbeta_{g_i^0}^0 \in \mR^p$ to be the {\it group specific} regression coefficient.
Let $\mC_g^0 = \{i:g_i^0=g\}$
{collect the indices of individuals in the $g$th group, and denote}
$\cG_{G_0}^0=\{\mC_1^0,\mC_2^0,\ldots,\mC_{G_0}^0\}$ as the corresponding partition
over $\{1,\ldots, N\}$.
{More generally, for any given integer $G\geq 1$,
we denote $\cG_G = \{\mC_1, \mC_2, \ldots, \mC_G\}$ as a partition of $\{1, \ldots, N\}$,
where $\mC_g$ collects the indices of individuals assigned to the $g$th group.
The associated group-specific coefficient vectors are denoted as $\bbeta = (\bbeta_1, \ldots, \bbeta_G)^\top \in \mathbb{R}^{G\times p}$.
For each individual $i$, we write $g_i \in \{1, \ldots, G\}$ for its group membership under partition $\cG_G$, therefore equivalently $i \in \mC_{g_i}$.}
In practice, the true group number $G_0$, partition $\cG_{G_0}^0$, and group parameters $\bbeta^0 = (\bbeta_1^0,\ldots, \bbeta_{G_0}^0)^\top\in\mathbb{R}^{G_0\times p}$ are all unknown
and need to be estimated.
To be more specific, given
a pre-specified group number $G$,
we estimate the unknown memberships and group parameters by minimizing the following loss function
$\mL(\bZ;\bbeta,\cG_G)$: 
\compressmath
\begin{align} \label{loss_function}
\big\{\wh \bbeta,\wh \cG_G \big\} = \argmin_{\bbeta,\cG_G}\mL(\bZ;\bbeta,\cG_G)=\argmin_{\bbeta,\cG_G}\frac{1}{NT} \sum_{i =1}^N\sum_{t = 1}^T\boldsymbol{\ell}(\bz_{it}; \bbeta_{g_i}),
\end{align}
where $\boldsymbol{\ell}(\z_{it};\bbeta_{g_i})$
represents the loss function evaluated at $\z_{it}$ with parameter $\bbeta_{g_i}$, $\wh\bbeta = (\wh\bbeta_1, \ldots, \wh\bbeta_G)^\top$ and $\wh\cG_G = \{\wh\mC_1, \ldots, \wh\mC_G\}$. 
%The model parameter {$\wh\bbeta = (\wh\bbeta_1, \ldots, \wh\bbeta_G)^\top$}
%and group partition {$\wh\cG_G = \{\wh\mC_1, \ldots, \wh\mC_G\}$} can be estimated by using a {$k$-means type algorithm} \citep{liu2020identification},
%which is given in Algorithm \ref{estimation_algorithm}.
The optimization in \eqref{loss_function} can be solved via a $k$-means type algorithm \citep{liu2020identification}, summarized in Algorithm \ref{estimation_algorithm}.

\begin{algorithm}[h]
\footnotesize
\setstretch{1.2}
\caption{A $k$-means Type Algorithm for Group Panel Models}
\label{estimation_algorithm}
\begin{algorithmic}[1]
\Require Number of groups $G$; dataset $\bZ$; loss function $\boldsymbol{\ell}(\cdot;\cdot)$;
         tolerance $\epsilon$; maximum number of iterations $S_{\max}$.
\Ensure  Estimators $\wh\bbeta$ and $\wh\cG_G$.
\State Initialize $\wh\bbeta^{(0)}=(\wh\bbeta_1^{(0)},\ldots,\wh\bbeta_G^{(0)})^\top\in\mR^{G\times p}$; set $s\gets 0$.
\Repeat
   \State \underline{\textbf{Update group membership.}} For $i=1,\ldots,N$,
   \[
      \wh g_i^{(s+1)} = \argmin_{g\in[G]} T^{-1}\sum_{t=1}^T \boldsymbol{\ell}\big(\bz_{it};\wh\bbeta_g^{(s)}\big).
   \]
   \Statex \qquad Set $\wh\mC_g^{(s+1)}=\{i:\wh g_i^{(s+1)}=g\}$ for $g\in[G]$, and $\wh\cG_G^{(s+1)}=\{\wh\mC_1^{(s+1)},\ldots,\wh\mC_G^{(s+1)}\}$.
   \State \underline{\textbf{Update coefficients.}} For $g=1,\ldots,G$,
   \[
      \wh\bbeta_g^{(s+1)} = \argmin_{\bbeta_g\in\mR^p}
      \sum_{i\in\wh\mC_g^{(s+1)}}\sum_{t=1}^T \boldsymbol{\ell}(\bz_{it};\bbeta_g).
   \]
   \State $s\gets s+1$.
\Until{$\wh g_i^{(s)}=\wh g_i^{(s-1)}$ for all $i$, or $\|\wh\bbeta^{(s)}-\wh\bbeta^{(s-1)}\|_F<\epsilon$, or $s\ge S_{\max}$.}
\State \Return $\wh\bbeta=\wh\bbeta^{(s)}$ and $\wh\cG_G=\wh\cG_G^{(s)}$.
\end{algorithmic}
\end{algorithm}

To estimate the unknown group number $G_0$, a popular way is to adopt the IC methods.
However, such methods still rely on a tuning constant in the penalty term that must be pre-specified \citep{su2016identifying,liu2020identification}.
%The selected group number $\wh G$ can be sensitive to the user-chosen constant, whereas the user-chosen constant may vary from the model and error distribution specifications, making it less robust in practical applications.
The choice of this constant may depend on model and error specifications, leading to instability in practice. 
Thus, we aim to develop a data-driven procedure to estimate $G_0$.

{\sc General Notation.}
{For a positive integer $n$, denote $[n] = \{1,\ldots, n\}$.}
 We use $\|\cdot\|_2$ to denote the Euclidean norm of a vector.
For any matrix $\M$, $\|\M\|_F = \sqrt{\text{tr}(\M^{\top}\M)}$ denotes the Frobenius norm of $\M$.
In addition, for any symmetric matrix $\M$, let $\lambda_{\max}(\M)$ and $\lambda_{\min}(\M)$ denote the maximum and minimum eigenvalues of $\M$, respectively.
Moreover, for any vector $\bv\in\mR^n$,
let $v_{[k]}$ denote its $k$th element and define $\|\bv\|_{\M} = (\bv^\top\M\bv)^{1/2}$ for any positive definite matrix $\M$.
Similarly, for any two vectors $\bv_1,\bv_2\in\mR^n$, define $\Xi_{\M}(\bv_1,\bv_2)=\bv_1^\top \M \bv_2$ for any positive definite matrix $\M$ and $\Xi(\bv_1,\bv_2)=\bv_1^\top \bv_2$.
For any twice-differentiable function $f(\bbeta)$ with respect to a vector $\bbeta$, we use $\dot f(\bbeta) = \partial f(\bbeta)/\partial \bbeta$ and $\ddot f(\bbeta) = \partial^2 f(\bbeta)/(\partial \bbeta\,\partial \bbeta^\top)$ to denote its first and second order derivatives, respectively.
For a sequence of positive real numbers $\{A_n\}$ and a sequence of random variables $\{X_n\}$, we denote $X_n\gtrsim A_n$ if there exists a constant $C>0$ such that $X_n \ge CA_n$  holds for large enough $n$ with probability tending to one.

\subsection{Quadratic Approximation to Objective Function}\label{subsec:loss_approx}

To estimate the number of groups in a data-driven manner, we need to construct a selection criterion with the cross-validation (CV) method.
Following a common practice in the literature \citep{yao1988estimating,zou2020consistent,zhu2021least}, the selection criterion is designed based on
a quadratic approximation to the loss function
$\boldsymbol{\ell}(\bz;\bbeta)$.
To motivate the idea, we first define the node-wise
loss function as
\begin{align}
\ol \mL_{i}(\bZ;\bbeta) = \frac{1}{T}\sum_{t = 1}^T
\boldsymbol{\ell}(\bz_{it}; \bbeta).\label{equ:node_wise_Li}
\end{align}
Subsequently, we could obtain a node-wise estimator by minimizing
$\ol \mL_{i}(\bZ;\bbeta)$ as
$\wh \bgamma_i = \argmin_{\bbeta}\ol \mL_{i}(\bZ;\bbeta)$.
Then we can approximate $\ol \mL_{i}(\bZ;\bbeta)$ by using a
Taylor expansion in the neighborhood of $\wh \bgamma_i$ as
\begin{align*}
\ol \mL_{i}(\bZ;\bbeta)& \approx \ol \mL_{i}(\bZ;\wh \bgamma_i)
+\frac{1}{2T}\sum_{t = 1}^T (\bbeta - \wh \bgamma_{i})^\top
\ddot {\boldsymbol{\ell}}(\bz_{it};\wh \bgamma_i) (\bbeta - \wh \bgamma_{i}),
\end{align*}
where the first term is not related to $\bbeta$.
Define $\wh \bW_i = T^{-1}\sum_{t=1}^T \ddot{\boldsymbol{\ell}}(\bz_{it};\wh \bgamma_i)$
and $\bs(\bz_{it};\bbeta) = \dot {\boldsymbol{\ell}}(\bz_{it}; \bbeta)$.
We note that $\wh \bW_i(\bbeta - \wh \bgamma_i) \approx T^{-1}\sum_{t=1}^T \bs(\bz_{it};\bbeta)$,
then we have
\compressmath
\begin{align}
\frac{1}{T}\sum_{t = 1}^T (\bbeta - \wh \bgamma_{i})^\top
\ddot{\boldsymbol{\ell}}(\bz_{it};\wh \bgamma_i) (\bbeta - \wh \bgamma_{i})
 &= (\bbeta - \wh \bgamma_{i})^\top
\wh \bW_i (\bbeta - \wh \bgamma_{i})\nonumber\\
& \approx \Big\{\frac{1}{T}\sum_{t=1}^T \bs(\bz_{it}; \bbeta)\Big\}^\top\wh \bW_i^{-1}\Big\{\frac{1}{T}\sum_{t=1}^T \bs(\bz_{it}; \bbeta)\Big\}.\label{local_quad_i}
\end{align}
For convenience,
we consider the following quadratic approximation to
the loss function in \eqref{loss_function} as
\vspace{-20pt}
\begin{align}
\mQ(\bZ;\bbeta, \cG_G)  = \frac{1}{N}\sum_{g =1}^G\sum_{i \in \mC_g}
\Big\{\frac{1}{T}\sum_{t=1}^T \bs(\bz_{it}; \bbeta_g)\Big\}^\top \wh\bW_i^{-1}\Big\{\frac{1}{T}\sum_{t=1}^T \bs(\bz_{it}; \bbeta_g)\Big\}\defeq  \frac{1}{N}\sum_{i=1}^N \ol\mQ_i(\bZ;\bbeta_{g_i})\label{LS_criterion}
\end{align}
by ignoring the constants.
We remark that the quadratic approximation in \eqref{LS_criterion} is valid provided the loss function $\boldsymbol{\ell}(\bz;\bbeta)$ possesses continuous second-order derivatives, which is satisfied by many commonly used models such as the linear panel model, probit panel model, and Poisson panel model.
\label{remark_Qfunc}

\subsection{Selection Criterion with Cross-Validation}\label{subsection:select_criterion_with_cv}

{Based on the quadratic approximation \eqref{LS_criterion}, we propose a selection criterion using 2-fold cross-validation.
With this partition, we split the data $\bZ$ into two folds $\Z^{(1)} = \{\z_{it}: i\in [N], t\in\mT^{(1)}\}$ and $\Z^{(2)} = \{\z_{it}: i\in [N], t\in\mT^{(2)}\}$ on the time span, as illustrated in Figure \ref{fig:split}.
To ensure that the dependence between the two data folds is negligible,
a buffer zone with time length $\tau_{NT}$ is specified between the two folds, i.e., $\mT^{(1)} = \{1,2,\ldots, \lfloor T/2 - \tau_{NT}/2\rfloor\}$ and $\mT^{(2)} =
\{\lfloor T/2 + \tau_{NT}/2\rfloor, \ldots, T\}$.
Such buffering (or gap) strategies have been widely adopted in cross-validation and sample-splitting procedures for dependent data \citep{racine2000consistent,bergmeir2018note,semenova2023inference}.
Our theoretical analysis suggests that $\tau_{NT}$ should satisfy $\tau_{NT}\gg \log N$ to guarantee that the dependence between the two data folds is asymptotically negligible under the $\beta$-mixing condition.
In practice, we recommend using $\tau_{NT} = (\log N)(\log T)^{\delta}$ with $\delta = 0.2$, which yields satisfactory numerical performance.

To conduct the CV method,
we first estimate the parameters and the group memberships using each data fold separately as
%\vspace{-14.9pt}
\begin{align*}
\qquad\{\wh \bbeta^{(k)}, \wh \cG_G^{(k)}\} = \argmin_{\bbeta, \cG_G} \mL(\bZ^{(k)};\bbeta, \cG_G),\quad k=1,2.
\end{align*}
Then we evaluate the {out-of-sample} loss on the other data fold, respectively as
$\mQ(\bZ^{(2)}; \wh \bbeta^{(1)}, \wh \cG_G^{(1)})$
and $\mQ(\bZ^{(1)}; \wh \bbeta^{(2)}, \wh \cG_G^{(2)})$.
Subsequently, we select $G$ by minimizing the following
 CV based criterion,
\vspace{-20pt}
\begin{align}
\wh G = {\argmin_{G\in[G_{\max}]}}\Big\{
\mQ(\bZ^{(2)}; \wh \bbeta^{(1)}, \wh \cG_G^{(1)}) +
\mQ(\bZ^{(1)}; \wh \bbeta^{(2)}, \wh \cG_G^{(2)})\Big\}\label{select_criterion},
\end{align}
{where $G_{\max }$ denotes the maximum number of groups under consideration.}
The selection procedure is summarized in Algorithm \ref{selection_algorithm}.
We remark that one can also consider a $K$-fold CV procedure for estimating $G$ to retain more training data, and the implementation is presented as follows.

\begin{algorithm}[H]
\footnotesize
\setstretch{1.2}
\caption{\footnotesize Group Number Selection for Group Panel Models}\label{selection_algorithm}
\begin{algorithmic}[1]
\Require
The maximum possible number of groups $G_{\max}$;
the dataset $\bZ$; buffer zone length $\tau_{NT}$.
\Ensure
Selected number of groups $\wh G$.
\State Split $\bZ$ into $\bZ^{(1)}$ and $\bZ^{(2)}$ with a buffer zone of length $\tau_{NT}$ as in Figure~\ref{fig:split}.
\For{$G=1,2,\ldots, G_{\max}$}
\State Use Algorithm \ref{estimation_algorithm} separately on $\bZ^{(1)}$ and $\bZ^{(2)}$   to obtain $\{\wh\bbeta^{(1)},\wh\cG_G^{(1)}\}$ and  $\{\wh\bbeta^{(2)},\wh\cG_G^{(2)}\}$.
\State Calculate the validation loss $\ol \mQ(G)\defeq \mQ(\bZ^{(2)}; \wh{\bbeta}^{(1)}, \wh \cG_G^{(1)})+\mQ(\bZ^{(1)}; \wh{\bbeta}^{(2)}, \wh \cG_G^{(2)}).$
\EndFor
\State \Return $\wh G = \argmin_{G\in[G_{\max}]} \ol\mQ(G)$.
\end{algorithmic}
\end{algorithm}

\begin{remark}
  {\sc ($K$-fold CV Implementation)}
  {First, we segment the time span $\mT = \{1,\ldots, T\}$ into $K$ consecutive folds of approximately equal length, where the time points in the $k$th fold are collected in the set $\wt \mT^{(k)}$.
Suppose that the $K$ data folds are segmented by the time points $t_1,\ldots, t_K$, where $t_k = \lfloor kT/K\rfloor$ for $1\le k\le K-1$ and $t_K = T$.
Then we specify the $k$th buffer zone as $\cB^{(k)} = \{\lfloor t_k -
\tau_{NT}/2\rfloor, \ldots, \lfloor t_k +
\tau_{NT}/2\rfloor\}$ and we set $\cB^{(0)} = \emptyset$ for convenience.
Subsequently, we evaluate the $\mQ$-criterion for $K$ rounds.
For the $k$th round, we take $\mT^{(k)}\defeq \wt\mT^{(k)}\backslash \{\cB^{(k)}\cup \cB^{(k-1)}\}$ as the time span in the testing data fold, and $\mT \backslash\{\wt\mT^{(k)}\cup \cB^{(k-1)}\cup \cB^{(k)}\}$ as the training data fold.
The above process is illustrated in Figure \ref{fig:splitK}.
Lastly, the group number is estimated by $\wh G = \argmin_G \sum_{k = 1}^K
\mQ(\Z^{(k)}; \wh \bbeta^{(-k)}, \wh \cG_G^{(-k)})$, where $\Z^{(k)}$ is the testing data fold in the $k$th round, and
$\{\wh \bbeta^{(-k)}, \wh \cG_G^{(-k)}\}$ denotes the estimates obtained {using the corresponding training fold.}
}
\end{remark}
\begin{figure}[htbp]
  \centering
  \includegraphics[scale=0.4]{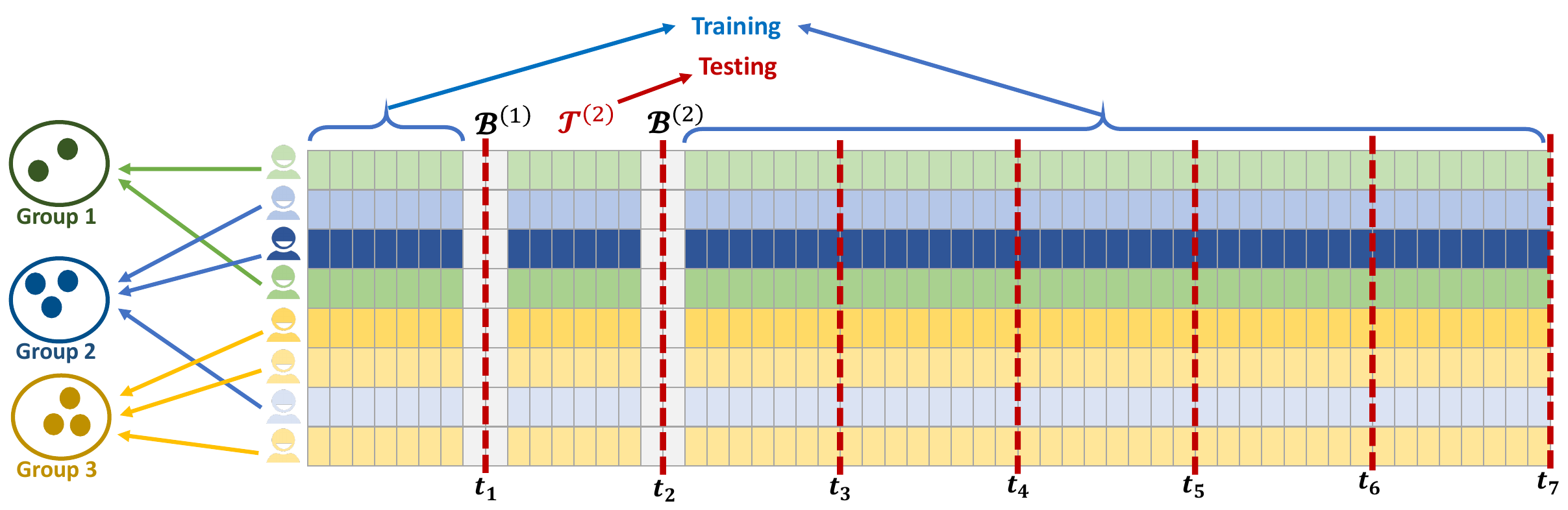}
  \caption{\small\it{Illustration of the $K$-fold cross-validation procedure with buffer zones for $K=7$, where $\mathcal{T}^{(2)}$ serves as the testing fold.}}\label{fig:splitK}
\end{figure}
We explain the rationale of \eqref{select_criterion} by decomposing
$ \mQ(\bZ^{(2)}; \wh\bbeta^{(1)}, \wh\cG_G^{(1)})$.
Specifically, we first
define $\ol \bs_i(\bZ^{(2)};\bbeta) = |\mT^{(2)}|^{-1}\sum_{t\in \mT^{(2)}}
\bs(\bz_{it};\bbeta)$
and $\bDelta_i(\bZ^{(2)};\bbeta_1, \bbeta_2) = \bar{\bs}_i(\bZ^{(2)};\bbeta_1) -\bar{\bs}_i(\bZ^{(2)};\bbeta_2 ) $ with any given $\bbeta, \bbeta_1,\bbeta_2$.
{Let $\wh \mC_g^{(1)}$ ($g = 1,\ldots, G$) denote the estimated group structure using $\bZ^{(1)}$. We can then decompose $\mQ(\bZ^{(2)}; \wh\bbeta^{(1)}, \wh\cG_G^{(1)})$ as}
\compressmath
\begin{align}
 \mQ(\bZ^{(2)}; \wh\bbeta^{(1)}, \wh\cG_G^{(1)}) & = \frac{1}{N}\sum_{g =1}^G\sum_{i \in \wh \mC_g^{(1)}}
\Big\|\bDelta_i(\bZ^{(2)}; \wh \bbeta^{(1)}, \wh \bbeta^{(2)})\Big\|_{\wh\bW_i^{-1}}^2+
\frac{1}{N}\sum_{g =1}^G\sum_{i \in \wh \mC_g^{(1)}}\Big\|\bar{\bs}_i(\bZ^{(2)}; \wh \bbeta^{(2)})\Big\|_{\wh\bW_i^{-1}}^2\nonumber\\
&~~~~~+\frac{2}{N}\sum_{g =1}^G\sum_{i \in \wh \mC_g^{(1)}}
\Xi_{\wh\bW_i^{-1}}\Big(\bDelta_i(\bZ^{(2)}; \wh \bbeta^{(1)}, \wh \bbeta^{(2)}),\bar{\bs}_i(\bZ^{(2)}; \wh \bbeta^{(2)})\Big)\nonumber\\
&\defeq {\small\mS\big(\bZ^{(2)}; \wh \bbeta^{(1)}, \wh \bbeta^{(2)}, \wh\cG_G^{(1)}\big)
+ \mathcal{D}\big(\bZ^{(2)}; \wh \bbeta^{(2)}, \wh\cG_G^{(1)}\big) +
2 \mathcal{R}\big(\bZ^{(2)}; \wh \bbeta^{(1)}, \wh \bbeta^{(2)}, \wh\cG_G^{(1)}\big),}\label{equ:err_test}
\end{align}
with the notation $\|\bv\|^2_\bW =\bv^\top\bW\bv$ and $\Xi_{\bW}(\bv_1, \bv_2) = \bv_1^\top \bW \bv_2$ for $\bv,\bv_1,\bv_2\in\mR^{p}$, $\bW\in\mR^{p\times p}$.
The third term is a cross term which is dominated by the first two terms under both underfitting and overfitting cases.
In the underfitting case (i.e., $G<G_0$),
$\mD(\bZ^{(2)}; \wh \bbeta^{(2)}, \wh\cG_G^{(1)})$
will dominate
with a large prediction error on the testing dataset.
In the overfitting case (i.e., $G>G_0$), we can show that
$\mS(\bZ^{(2)}; \wh \bbeta^{(1)}, \wh \bbeta^{(2)}, \wh\cG_G^{(1)})$ will dominate
by taking consideration of the optimization procedure on the training data fold.
Consequently,
$\mS(\bZ^{(2)}; \wh \bbeta^{(1)}, \wh \bbeta^{(2)}, \wh\cG_G^{(1)})$ plays a role similar to the penalty term in the BIC or AIC selection, while
using a data-driven strategy.
This avoids setting a user-determined constant in information criteria.
\begin{remark}
We consider a simple univariate model for illustration, i.e.,
$
Z_{it} = \beta_{g_i} + \ve_{it},
$
where $\ve_{it}$ is the independent noise term with mean 0 and variance $1$.
For estimation, we use the least squares objective function, i.e.,
$N^{-1}|\mT^{(1)}|^{-1}\sum_{i = 1}^N \sum_{t\in \mT^{(1)}}
(Z_{it} - \beta_{g_i})^2$.
Suppose the group structure is estimated using the first fold $\bZ^{(1)}$ as
$\wh \mC_g^{(1)}$ for $g\in[G]$.
Then we have $\wh \beta_g^{(1)} = |\wh \mC_g^{(1)}|^{-1}|\mT^{(1)}|^{-1}\sum_{i\in \wh \mC_g^{(1)}}
\sum_{t\in \mT^{(1)}}Z_{it}$ and
$\wh \beta_g^{(2)} = |\wh \mC_g^{(1)}|^{-1}|\mT^{(2)}|^{-1}\sum_{i\in \wh \mC_g^{(1)}}
\sum_{t\in \mT^{(2)}}Z_{it}$.
In this case, we have
$\ol s_i(\bZ^{(2)};\beta) = |\mT^{(2)}|^{-1}
\sum_{t\in \mT^{(2)}}(\beta -  Z_{it})=\beta-|\mT^{(2)}|^{-1}\sum_{t\in \mT^{(2)}}Z_{it}$ and
$\Delta_i(\bZ^{(2)}; \wh \beta_g^{(1)}, \wh \beta_g^{(2)}) = \ol s_i(\bZ^{(2)};\wh \beta_g^{(1)})-\ol s_i(\bZ^{(2)};\wh \beta_g^{(2)})=\wh \beta_g^{(1)} - \wh \beta_g^{(2)}
$.
This yields
\beq
\mQ(\bZ^{(2)}; \wh \bbeta^{(1)}, \wh\cG_G^{(1)})  =
 \frac{1}{N}\sum_{g = 1}^G\sum_{i\in \wh \mC_g^{(1)}}
\Big(\frac{1}{|\mT^{(2)}|}\sum_{t\in \mT^{(2)}}Z_{it} - \wh\beta_g^{(2)}\Big)^2+\frac{1}{N}\sum_{g = 1}^G |\wh \mC_g^{(1)}| (\wh \beta_g^{(1)} - \wh \beta_g^{(2)})^2,\label{Q_func_simple}
\eeq
and
$\mathcal{R}(\bZ^{(2)}; \wh \bbeta^{(1)}, \wh \bbeta^{(2)}, \wh\cG_G^{(1)})= 0$ in this case since $\sum_{i\in\wh \mC_g^{(1)}}\ol s_i(\bZ^{(2)};\wh\beta_g^{(2)})=0$.
The form is similar to the IC based method for determining the group number \citep{bonhomme2015grouped,su2016identifying,liu2020identification}.
The first term in \eqref{Q_func_simple} is used to evaluate the goodness-of-fit level, and the second term plays the role of penalty for model complexity as in the IC method.
In contrast to the user-specified tuning parameter typically involved in the second penalty term for the IC method, the amount of ``penalty'' in \eqref{Q_func_simple} can be totally determined by the data information.
As a consequence, our data-splitting method is tuning-free and thus more user-friendly.
\end{remark}

\subsection{Comparison with Existing Methods}

In this section, we compare the proposed CV method with the widely used information criterion (IC) and hypothesis testing (HT) methods in detail.
In addition, we also compare with the CV method for the change points' number estimation problem to clarify the differences.

\subsubsection{Comparison with IC and HT Methods}\label{sec:comp_IC}

We first compare the proposed CV based estimation with two widely used methods for estimating $G$ in the group panel data literature.
The first is the IC based methods \citep{bonhomme2015grouped,su2016identifying,liu2020identification}, and the second is the hypothesis testing-based method \citep{lu2017determining}.
The IC based methods minimize the criterion in the following form, i.e.,
\compressmath
\begin{align}
\IC(G) = \frac{1}{NT} \sum_{i =1}^N\sum_{t = 1}^T\boldsymbol{\ell}(\bz_{it}; \wh \bbeta_{\wh g_i}) + \rho_{\lambda_{NT}}(G),\label{eq:IC}
\end{align}
where $\rho_{\lambda_{NT}}(G)$ is the penalty function with model-specific tuning parameter $\lambda_{NT}$.
Particularly, for different panel data models, different $\rho_{\lambda_{NT}}(G)$ functions are utilized.
For the linear panel data model,  \cite{bonhomme2015grouped} used a BIC based criterion and set
$\rho_{\lambda_{NT}}(G) = \wh \sigma^2({NT})^{-1}({GT+N+p})\log(NT)$, where
$\wh \sigma^2$ is an estimated variance by setting $G$ as a sufficiently large value.
However, they did not provide the theoretical analysis for estimating $G$.
Subsequently, \cite{su2016identifying} and \cite{liu2020identification} studied more generalized panel data models but they set different penalty forms for different models.
To be more specific,
\cite{su2016identifying} recommended using $\rho_{\lambda_{NT}}(G) = p\lambda_{NT}G$ with $\lambda_{NT} = 1/3(NT)^{-1/2}$  for the linear  model,
and $\lambda_{NT} = \log(\log T)/(8T)$
 for the probit model.
In contrast, \cite{liu2020identification} set $\rho_{\lambda_{NT}}(G) = \lambda_{NT}G$ with  $\lambda_{NT} = 1/\{5\log(T)T^{1/8}\}$  for the linear  model,
and $\lambda_{NT} = \log(N)^{1/8}/\{5(\log T)T^{1/8}\}$ for the probit model.
We summarize different penalty functions in the IC methods in Table \ref{tab:penalty_functions}.
In addition to the IC methods, the  hypothesis testing method was studied by
\cite{lu2017determining} for the linear panel data model.
They considered the testing problem as
\compressmath
\begin{align*}
	\mathbb{H}_{0}\left(G_{0}\right): G=G_{0} \quad \operatorname{versus} \quad \mathbb{H}_{1}\left(G_{0}\right): G_{0}<G \leq G_{\max},
\end{align*}
and they derived a residual-based LM-type test statistic.
However, their test statistic can only be applied to linear models.
%\vspace{-5mm}
\begin{table}[htbp]
\centering
\caption{Summary of penalty functions in different IC methods.}
\begin{tabular}{lccc}
\toprule
 & \textbf{Model} & ${\rho_{\lambda_{NT}}(G)}$ & ${\lambda_{NT}}$ \\
\midrule
\multirow{2}{*}{\cite{bonhomme2015grouped}} & \multirow{2}{*}{Linear} & \multirow{2}{*}{$  \frac{({GT+N+p})\log(NT)}{NT}\cdot\wh\sigma^2$} & \multirow{2}{*}{--} \\
~\\ \hline
\multirow{2}{*}{\cite{su2016identifying}} & Linear & $p\lambda_{NT}G$ & $(NT)^{-1/2}/3$ \\
& Probit & $p\lambda_{NT}G$ & ${\log(\log T)}/{(8T)}$ \\
\hline
\multirow{2}{*}{\cite{liu2020identification} } & Linear & $\lambda_{NT}G$ & $\{5\log(T)T^{1/8}\}^{-1}$ \\
& Probit & $\lambda_{NT}G$ & ${\log(N)^{1/8}}\{5(\log T)T^{1/8}\}^{-1}$ \\
\bottomrule
\end{tabular}
\label{tab:penalty_functions}
\end{table}
%\vspace{-5mm}
In summary, compared with these existing methods, we highlight the merits of our CV based method as follows.
First, we provide a unified group number selection method, which can be flexibly applied to various panel data models.
In contrast, the IC based method requires specifying the penalty function \eqref{eq:IC} individually for each panel data model, while the hypothesis testing method needs to derive valid testing rules case by case.
Actually, as shown by our simulation study,
we find that our CV method can work well under {various panel models}, while the IC methods \citep{su2016identifying,liu2020identification} generally yield larger estimation errors
when applied beyond the specific models for which their penalties were tuned.
Second, the proposed CV method is model-specific tuning free, namely, it does not need to select extra tuning parameters ($\lambda_{NT}$) as the IC methods.
The selection of $\lambda_{NT}$ can be subjective and may not be adaptive to
various datasets.
%As shown by our numerical studies, we find that the CV method is able to achieve better performances when the group differences are smaller (i.e., weak signals), which indicates its  robust performance.
Third, our numerical studies show that the CV method achieves better performance under weak signals (i.e., when the group differences are small), demonstrating its robustness.

\subsubsection{Comparison with the CV Method in Change Points Number Estimation}\label{sec:comp_Zou}

The proposed method shares great similarity with the CV method
designed by \cite{zou2020consistent} for estimating the number of change points.
They consider a similar quadratic  loss function for estimating the number of change points.
Despite the similarity, we would like to highlight two differences between the proposed methodology with their approach.
First, the data structure is different for the change point detection problem and panel data modeling.
To be more specific, \cite{zou2020consistent} considered a one dimensional problem, i.e., $\{Z_t: 1\le t\le T\}$, and aimed to estimate the change points along $1\le t\le T$.
{They split the time points into odd- and even-indexed subsets to preserve the time dynamic patterns on each data fold when conducting the CV method.}
On the other hand, the panel data $\{\z_{it}: 1\le i\le N, 1\le t\le T\}$ contains two dimensions: the individuals ($\{i\in [N]\}$) and the time span ($\{t\in [T]\}$).
This leads to different definitions of ``groups'' under the above two settings.
In the change points detection problem, {\it groups} are defined over the time span to be segmentations of $1\le t\le T$.
In contrast, in the panel data setting, {\it groups} refer to partitions of individuals $\{i: 1\le i\le N\}$.
{%As a result, it leads to different optimization algorithms.
For the group panel data model, we typically use a $k$-means type algorithm to partition among the individuals,
while for change point detection, the binary segmentation and optimal partitioning algorithms are utilized for identifying the segmentations along the time span.}

Second, in the panel data setting, determining how to partition the dataset is crucial, given its two-dimensional structure.
The first splitting scheme is to split the dataset on the individual dimension, {which is referred to as individual-based splitting}. In this way, the time dynamics pattern can be preserved on each data fold.
The second  is to split on the time dimension, {which is referred to as temporal-based splitting}. In this way, the group structure of all individuals can be preserved.
{For comparison, we conduct the CV methods for both schemes with a simple simulation example, where the details are stated in Appendix \ref{subsec:split_N} and we include Figure \ref{fig:split_NT} here for illustration.}
\begin{figure}[htbp]
  \centering
  \includegraphics[scale=0.4]{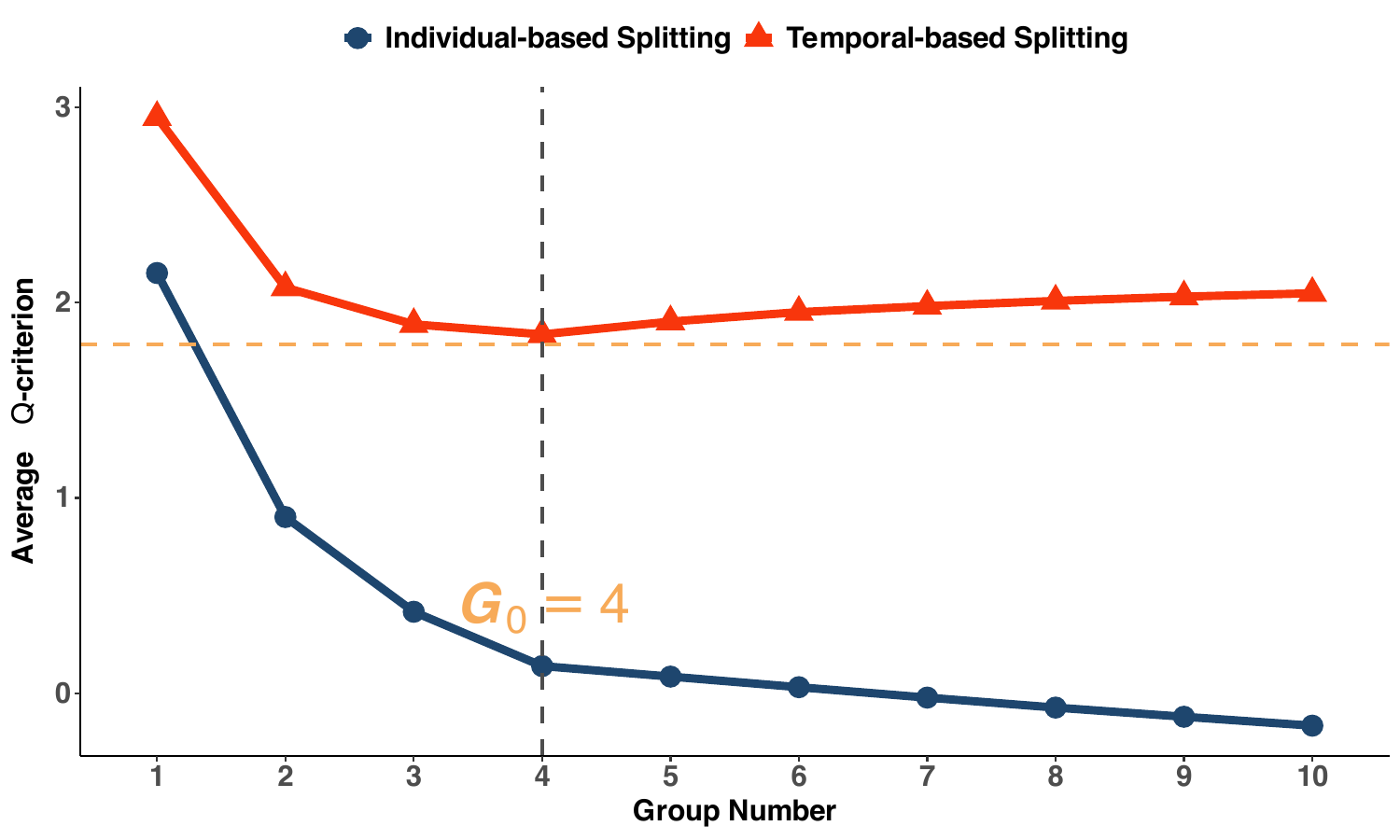}
  \caption{\small\it {Average $\mQ$-criterion} for the two splitting schemes. The blue curve represents individual-based splitting, and the red curve represents temporal-based splitting.  The annotation $G_0=4$ indicates the true number of groups.}\label{fig:split_NT}
\end{figure}
\vspace{-5mm}
It shows that the $\mQ$-criterion \eqref{LS_criterion} tends to monotonically
{decrease} as $G$ increases for the individual-based splitting scheme, while it exhibits an elbow point at $G_0$ for the temporal-based splitting scheme.
This implies the temporal-based splitting scheme (adopted in our framework) is the appropriate choice.
We further discuss the intuitions behind this phenomenon in Remark \ref{rem:split}.

\section{Theoretical Properties}\label{sec:theoretical_properties}

\subsection{Technical Conditions}\label{sec:conditions}

We first introduce multi-index notation to handle higher-order partial derivatives in a unified manner. 
For $\bbeta=(\beta_{[1]},\ldots,\beta_{[p]})^\top\in\mR^p$ and a multi-index 
$\boldsymbol{m}=(m_1,\ldots,m_p)$ with non-negative integer entries, denote 
$|\boldsymbol{m}|=\sum_{l=1}^p m_l$ and define the $|\boldsymbol{m}|$th order partial derivative
\begin{align*}
	D^{\boldsymbol{m}}\boldsymbol{\ell}(\bz;\bbeta)\ :=\ 
	\frac{\partial^{|\boldsymbol{m}|}\boldsymbol{\ell}(\bz;\bbeta)}
	{\partial \beta_{[1]}^{m_1}\,\partial \beta_{[2]}^{m_2}\cdots \partial \beta_{[p]}^{m_p}}.
\end{align*}
To establish the theoretical properties, the following conditions are required.

\makeatletter
\renewcommand{\theenumi}{\@arabic\c@enumi}
\renewcommand{\labelenumi}{(C\theenumi)}
\renewcommand{\theenumii}{\arabic{enumii}}
\renewcommand{\labelenumii}{(C\theenumi.\theenumii)}
\renewcommand{\p@enumi}{C}
\renewcommand{\p@enumii}{C\arabic{enumi}.}
\makeatother
\begin{enumerate}

\item (\textsc{Parameter Space}) Suppose there exists a constant $R>0$ such that\label{condition:param_space}
$
    \max_{g\in [G]} \|\bbeta_g\|_2\le R.
$

\item (\textsc{Distribution})
The observations satisfy the following dependence structure.
\label{condition:distribution}
\begin{enumerate}
\item (\textsc{Time Dependence}) For each $i \in [N]$, the process $\{\bz_{it}:t\in [T]\}$ is stationary and $\beta$-mixing with mixing coefficients $\beta_{i}(\cdot)$.
    \label{condition:stationary}
Moreover, $\beta(\tau) \defeq \sup _{N \geq 1} \max _{i\in[N]} \beta_{i}(\tau)$ satisfies $\beta(\tau) \leq 2\exp \left(-C_{0} \tau^{b_{0}}\right)$ for all $\tau \geq 0$ and some positive constants $C_{0}$ and $ b_{0}\ge 1$.
\item (\textsc{Individual Independence})  $\{\bz_{it}, t \in [T]\}$ are independent across $i\in[N]$.\label{condition:independency}
\end{enumerate}

\item (\textsc{Buffer Length}) The length of the buffer zone $\tau_{NT}$ between the two folds satisfies
$\tau_{NT}\to\infty$, $\tau_{NT}/\log(N)\to\infty$, and $\tau_{NT}=o(T)$ as $N,T\to\infty$.\label{condition:buffer_len}

{\item (\textsc{Smoothness})\label{condition:exp_tail}
There exists a non-negative function $K(\bz_{it})$ such that, for all
$\bbeta,\bbeta'\in\mR^p$ with $\|\bbeta\|_2\le R$ and $\|\bbeta'\|_2\le R$, and for every multi-index $\boldsymbol{m}$ with $|\boldsymbol{m}|\le 2$,
\begin{align}
	|D^{\boldsymbol{m}}\boldsymbol{\ell}(\bz_{it};\bbeta) - D^{\boldsymbol{m}}\boldsymbol{\ell}(\bz_{it};\bbeta')|
	\ \le\ K(\bz_{it})\,\|\bbeta-\bbeta'\|_2,
	\label{L_diff}
\end{align}
and $|D^{\boldsymbol{m}}\boldsymbol{\ell}(\bz_{it};\bbeta)|\le K(\bz_{it})$.
The function $K$ satisfies the exponential tail condition
\begin{align}
	\sup_{N\ge 1}\sup_{i\in[N]} P\!\left(K(\bz_{it})>v\right)
	\ \le\ \exp\!\left\{1-(v/B_1)^{b_1}\right\},
	\qquad \text{for all } v>0, \label{K_exp}
\end{align}
for some constants $b_1>0$ and $B_1>0$.
Furthermore, there exists a function $M(\bz_{it})$ with $E\{M(\bz_{it})\}<\infty$ 
such that $|D^{\boldsymbol{m}}\boldsymbol{\ell}(\bz_{it};\bbeta)|\le M(\bz_{it})$ 
for every multi-index $\boldsymbol{m}$ with $|\boldsymbol{m}|=3$ and every $\bbeta$ with $\|\bbeta\|_2\le R$.
\item (\textsc{Bounded Moments})\label{condition:higher_order_smoothness}
There exist a non-negative function $M^*(\bz_{it})$ and two integers $q_1$, $q_2$ with $1\le q_1\le 4$ and $q_2\ge 4$ such that, for every multi-index $\boldsymbol{m}$ with $|\boldsymbol{m}|=q_1$ and every $\bbeta$ with $\|\bbeta\|_2\le R$, the following holds almost surely,
\begin{align}
	|D^{\boldsymbol{m}}\boldsymbol{\ell}(\bz_{it};\bbeta)|\ \le\ M^*(\bz_{it})
	\quad \text{and}\quad
	\sup_{i\in[N]} E\!\left[\{M^*(\bz_{it})\}^{q_2}\right]<\infty.
\end{align}}
\vspace{-1cm}
\item (\textsc{Convexity})\label{condition:convex}
For each $g\in[G_0]$, define $\bH_g^\dag(\bbeta) = E\{\ddot{\boldsymbol{\ell}}(\bz_{it};\bbeta)\}$ for $i\in\mC_g^0$. There exist positive constants $c_0$ and $c_1$ such that $c_0\ \le\ \min_{g\in[G_0]} \lambda_{\min}\!\big(\bH_g^\dag(\bbeta_g^0)\big)
	\ \le\ \max_{g\in[G_0]} \lambda_{\max}\!\big(\bH_g^\dag(\bbeta_g^0)\big)\ \le\ c_1$.

\item (\textsc{Identification}) \label{condition:identification}
Let $\mL_i^*(\bbeta)=E\{\ol\mL_i(\bZ;\bbeta)\}$.
The following identification condition holds: for every $\epsilon>0$,
\begin{align*}
	\phi(\epsilon)\ :=\ \inf _{N \geq 1}\, \inf _{i\in[N]}\, 
	\inf _{\|\bbeta-\bbeta_{g_{i}^0}^{0}\|_{2}^2 \ge \epsilon}
	\left[\mL_{i}^*(\bbeta)-\mL_{i}^*(\bbeta_{g_i^0}^0)\right]\ >\ 0.
\end{align*}
The same condition also holds for $\mQ_i^*(\bbeta)=E\{\|\bar\bs_i(\bZ;\bbeta)\|_{\bW_i^{-1}}^2\}$, where $\bW_i=E(\ddot{\boldsymbol{\ell}}(\bz_{it};\bbeta_{g_i^0}^0))$.

\item (\textsc{Sample Size}) Let $d = b_0 b_1 / (b_0 + b_1)$ with $d \in (0,1)$. Suppose $N, T \to \infty$ and for some $\delta > 0$, $\log N = o\!\left( T^{\alpha - \delta} \right)$ with $\alpha=\min\{1/3,\, d/(2-d)\}$.
\label{condition:sample_size}

\item (\textsc{Group Size})  For all $g \in [G_0]$, there exists a positive constant $\pi_g$ such that $N_g/N \rightarrow \pi_g$ as $N \rightarrow \infty$. \label{condition:group_size}

\item (\textsc{Group Separation}) Let $d_{0}\defeq \min_{g^{\prime} \neq g}\left\|\bbeta_{g^{\prime}}^0-\bbeta_{g}^0\right\|_{2}>0$.\label{condition:group_separate}

\end{enumerate}

We explain the rationale of the technical conditions in detail as follows.
First, Condition \eqref{condition:param_space} assumes that the parameter space is compact, which is a regular condition in literature \citep{bonhomme2015grouped,su2016identifying,liu2020identification}.
Condition \eqref{condition:distribution} specifies the dependence structure of the observations.
Specifically, \eqref{condition:stationary} assumes that for each individual,
weak time dependence is allowed across the time dimension, which is milder than the {\it i.i.d.} condition in \cite{zou2020consistent}.
\eqref{condition:independency} assumes cross-sectional independence, i.e., the individuals are mutually independent. 
This dependence assumption is widely specified in the panel data literature \citep{gao2007nonlinear,bonhomme2015grouped,sarafidis2015partially,liu2020identification}. 
Condition \eqref{condition:buffer_len} regulates the length of the buffer zone introduced in our cross-validation scheme to mitigate the temporal dependence between the two data folds \citep{semenova2023inference}. 
Condition \eqref{condition:exp_tail} assumes a certain extent of smoothness for the loss function as well as its derivatives.
In addition, the exponential bound is imposed on the tail probability of the related function, and similar conditions are assumed by \cite{bonhomme2015grouped} and \cite{liu2020identification}.
Condition \eqref{condition:higher_order_smoothness} basically requires a bounded moment condition for the $q_2$th moment
of the derivatives, which is also assumed in the relevant literature
to ensure the grouping accuracy \citep{hahn2004jackknife,su2016identifying,liu2020identification}.

Next, Condition \eqref{condition:convex} assumes convexity within the interested area, and
Condition \eqref{condition:identification} is an identification condition imposed on each individual to ensure the local minimum at the true value.
\label{l4p:NT_explain}Condition \eqref{condition:sample_size}
allows $N$ to grow exponentially fast as $T\to\infty$.
As a consequence, it is suitable for panel data with large $N$  and moderately large $T$.
We remark that this condition is milder than the assumption imposed in several existing methods \citep{hahn2004jackknife,su2016identifying}, which require $N$ should not grow  polynomially fast with
$T$.
In addition, we remark that we require $N\to \infty$ to extract the major term
under the overfitting scenario, which is further shown to be lower bounded in Theorem \ref{thm:consistency2}.
Condition \eqref{condition:group_size} assumes that the group size $N_g$ should diverge at the same rate as the total number of individuals $N$.
Lastly, Condition \eqref{condition:group_separate} imposes a sufficiently large gap between the true group coefficients,
which is necessary for group membership estimation.
By Conditions \eqref{condition:param_space}--\eqref{condition:group_separate}, we can obtain the group estimation consistency result when $G\ge G_0$ \cite[Theorem 2]{liu2020identification}.
 Specifically, for each estimated group $\wh{\mC}_g=\{i: \wh{g}_i = g\}$, the group estimation consistency implies that there exists a true group $\mC_{\tilde g}^0 = \{i: g_i^0 = \wt g\}$ such that $\lim_{\min(N,T)\rightarrow \infty}P(\wh{\mC}_{g} \subseteq \mC_{\wt g}^0 )=1$.

\subsection{Selection Consistency}\label{subsec:selection_consistency}

Recall that $\cG_{G_0}^0 = \{\mC_1^0,\mC_2^0,\ldots, \mC_{G_0}^0\}$ is the true group partition with
$\mC_g^0 = \{i: g_i^0 = g\}$, and we treat $\cG_{G_0}^0$ as fixed memberships without randomness.}
Similarly, 
$\wh{\cG}_{G}^{(k)} = \{\wh{\mC}_1^{(k)},\wh{\mC}_2^{(k)},\ldots, \wh{\mC}_{G}^{(k)}\}$
is the estimated group partition using $\Z^{(k)}$
when $G$ groups are specified,
where $
\wh{\mC}_g^{(k)} = \{i: \wh{g}_i^{(k)} = g\}$.
{In the following we aim to prove the estimation consistency of the group number $G_0$ using the proposed CV estimation procedure.
It suffices to show that $\mQ(\bZ^{(2)}; \wh \bbeta^{(1)}, \wh \cG_G^{(1)})>
\mQ(\bZ^{(2)}; \wh \bbeta^{(1)}, \wh \cG_{G_0}^{(1)})$ holds with probability tending to one under both the underfitting case (i.e., $G<G_0$)
and the overfitting case (i.e., $G>G_0$).
For the underfitting part,
the proof  basically relies on the group difference condition \eqref{condition:group_separate}.
We show that there exists at least one group $g$ that is sufficiently distant from the others, leading to a large $\mQ$ value when $G<G_0$.

However, the proof for the overfitting part is more challenging and we illustrate the basic idea as follows.
Through a careful decomposition of $
\mQ(\bZ^{(2)}; \wh \bbeta^{(1)}, \wh \cG_{G_0}^{(1)})-
\mQ(\bZ^{(2)}; \wh \bbeta^{(1)}, \wh \cG_G^{(1)})$ in our theoretical analysis, we extract the leading term as
\begin{align}
  \mQ(\bZ^{(2)}; \wh \bbeta^{(1)}, \wh \cG_{G_0}^{(1)})-
\mQ(\bZ^{(2)}; \wh \bbeta^{(1)}, \wh \cG_G^{(1)})  = \mQ(\bZ^{(1)};\wh{\bbeta}^{(1)},{\cG_{G_0}^0}) -
	\mQ(\bZ^{(1)};\wh{\bbeta}^{(1)},\wh\cG_{G}^{(1)}) + o_p(T^{-1}).\label{eq:dom}
\end{align}
The establishment of the above result relies on our $(N,T)$ condition in \eqref{condition:sample_size}.
This interesting observation links the prediction loss on $\Z^{(2)}$
with the training loss reduction using $\Z^{(1)}$.
Similar results are also established by \cite{zou2020consistent} when estimating the number of change points with CV method,  but we rely on different techniques to show this result.
As long as $\mQ(\bZ^{(1)};\wh{\bbeta}^{(1)},{\cG_{G_0}^0}) -
	\mQ(\bZ^{(1)};\wh{\bbeta}^{(1)},\wh\cG_{G}^{(1)})\gtrsim T^{-1}$, the estimation consistency can be proved.
We formally state this  result in the following Theorem \ref{thm:consistency1}.

}

\bet\label{thm:consistency1}
Suppose conditions \eqref{condition:param_space}-\eqref{condition:group_separate} hold.
Further assume for $G > G_0$,
\compressmath
\begin{align}
\mQ(\bZ^{(k)};\wh{\bbeta}^{(k)},\cG_{G_0}^0) -
	\mQ(\bZ^{(k)};\wh{\bbeta}^{(k)},\wh\cG_{G}^{(k)}) \gtrsim T^{-1},
\label{lower_bound_Q}
\end{align}
for $k = 1,2$ with probability  tending to one.
Then we have $\lim_{\min(N,T)\rightarrow \infty} P(\wh G = G_0) = 1$.
\eet

The proof of Theorem \ref{thm:consistency1} is provided in Appendix \ref{subsection:proof_of_thm_consistency1}.
{It implies that the estimation consistency can be obtained as long as
the $\mQ$-criterion continues to decrease sufficiently as $G$ increases on the training data fold.
We then aim to establish the above critical lower bound condition \eqref{lower_bound_Q}.
First, under the assumed conditions, we show that $\mQ$ difference can be sufficiently approximated by the loss difference, i.e.,
$\mL(\bZ^{(k)};\wh{\bbeta}^{(k)},\cG_{G_0}^0) -
	\mL(\bZ^{(k)};\wh{\bbeta}^{(k)},\wh\cG_{G}^{(k)})$.
This approximation error is negligible with certain estimation precision for $\{\wh \bbeta_{\wh g_i}\}$ (see Lemma \ref{lem:LQ_diff} for details).
We note that $\mL(\bZ^{(k)};\wh{\bbeta}^{(k)},\wh\cG_{G}^{(k)})$ is minimized by exhausting all $G$-partitions of $[N]$, thus it should be smaller than
the optimal partition that partitions $\cG_{G_0}^0$ further into $G$ groups.
We denote the {corresponding loss} function by $\mL(\bZ^{(k)};\wh{\bbeta}^{(k)},\wh\cG_{G_0\to G}^{(k)})$.
This allows us to lower bound the left side of \eqref{lower_bound_Q} by
$\mL(\bZ^{(k)};\wh{\bbeta}^{(k)},\cG_{G_0}^0) -
	\mL(\bZ^{(k)};\wh{\bbeta}^{(k)},\wh\cG_{G_0\to G}^{(k)})$.
With this insight, we establish the lower bound in the following theorem.

}

\bet\label{thm:consistency2}
Suppose conditions \eqref{condition:param_space}--\eqref{condition:group_separate} hold. Then the lower bound \eqref{lower_bound_Q} is valid.

\eet

The proof of Theorem \ref{thm:consistency2} is provided in Appendix \ref{subsection:proof_of_thm_consistency2}.
{We explain the rationale as follows.
As we comment before, the lower bound condition \eqref{lower_bound_Q} is further reduced to
investigating the optimal partition of the true groups.
To illustrate our basic idea, we consider the case when $G$ is increased from $G_0$ to $G_0+1$, and thus
there exists one true group $\mC_g^0$ partitioned into $\mC_g^1$ and $\mC_g^2$.
Define
$$Q(\bZ;\bbeta,\mC_g)=\frac{1}{N_g}\sum_{i \in \mathcal{C}_{g}} \ol\bs_i\left(\Z ; \bbeta\right)^{\top} {\wh\bW_{i}^{-1}}\ol\bs_i\left(\Z ; \bbeta\right).
$$
It can be shown that $\mL(\bZ^{(k)};\wh{\bbeta}^{(k)},\cG_{G_0}^0) -
	\mL(\bZ^{(k)};\wh{\bbeta}^{(k)},\wh\cG_{G_0\to G}^{(k)})$ can be approximated by
\begin{align}
\frac{1}{N}\max_{\mC_g^1, \mC_g^2}
\Big\{
N_g Q(\bZ;\wh\bbeta_g,\mC_g^0) - N_{g_1} Q(\bZ;
 \wh \bbeta_{g,k},\mC_g^1)-N_{g_2}Q(\bZ;\wh \bbeta_{g,k},\mC_g^2)\Big\},\label{eq:train_reduce}
\end{align}
where $\wh \bbeta_{g,k}$ is the estimates obtained with individuals in  $\mC_g^k$ for $k = 1,2$, and
$N_{g_k} = |\mC_g^k|$.
Further analysis shows that \eqref{eq:train_reduce} can be simplified as
$\max_{\mC_g^1, \mC_g^2}\Delta(\mC_g^1, \mC_g^2)+o_p(T^{-1})$ with
\begin{align}
\Delta(\mC_g^1, \mC_g^2) = \frac{1}{N}\frac{N_{g_1}N_{g_2}}{N_g}
\Big\|\ol{\ol \s}_{\mC_g^1}
-\ol{\ol \s}_{\mC_g^2}\Big\|_{\bW^{(g)^{-1}}}^2,\label{eq:Delta_g1g2}
\end{align}
where $\ol{\ol \s}_{\mC_g^k} = N_{g_k}^{-1}\sum_{i\in\mC_g^k}\bar\bs_i(\bZ;\bbeta_g^0)$, and
 $\W^{(g)} = E(\wh \W_i)$ for $i\in \mC_g^0$.
As a result, it aims to find a partition $\{\mC_g^1, \mC_g^2\}$ which maximizes the square of  differences of two sample means, i.e., $\ol{\ol \s}_{\mC_g^1}$
and $\ol{\ol \s}_{\mC_g^2}$.
We visualize this procedure in Figure \ref{fig:max_Cg0} for illustration.
Note that as $T\to \infty$, we should have $\ol \s_i(\Z; \bbeta_g^0)$ asymptotically converges to a normal distribution.
This simplification allows us to derive the lower bound for $\max_{\mC_g^1, \mC_g^2}\Delta(\mC_g^1, \mC_g^2)$; see Lemma \ref{lem:U_inf_K} for details.

\begin{figure}[htbp]
  \centering
  \includegraphics[scale=0.5]{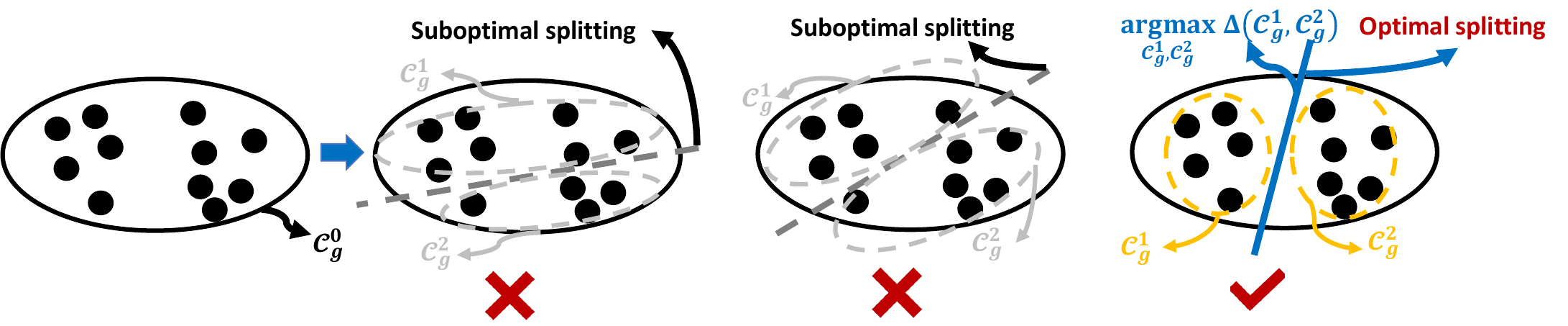}
\caption{\small\it{The solid blue line represents the optimal splitting that maximizes $\Delta(\mC_g^1, \mC_g^2)$; the gray dashed lines show two suboptimal splits for comparison.}}\label{fig:max_Cg0}
\end{figure}

Lastly, we comment on the possible gap
between $\min_{\bbeta, \cG_G}\mL(\bZ;\bbeta,\cG_G)$ and the minimum attained by the algorithm.
To minimize the loss function, we typically utilize a $k$-means type algorithm for model estimation; see Algorithm \ref{estimation_algorithm}.
Usually, a global optimum cannot be guaranteed by the $k$-means algorithm, but a local minimum can be assured.
To partially avoid this issue, we follow \cite{liu2020identification} to specify multiple initial values
to ensure a good convergence, and it yields satisfactory numerical performance in our simulation studies.
According to our theoretical analysis, the conclusion in Theorem \ref{thm:consistency2} still holds if
the minimum achieved by the algorithm stays not too distant from $\min_{\bbeta, \cG}\mL(\bZ;\bbeta,\cG_G)$.
We present the requirement in the following corollary.
\begin{corollary}\label{corollary_algo_gap}
	Let
$\mL_{\alg}(\bZ^{(k)};\wh{\bbeta}_{\alg}^{(k)},\wh\cG_{G,{\alg}}^{(k)})$ denote the minimum loss attained by the algorithm, and assume $\mL_{\alg}(\bZ^{(k)};\wh{\bbeta}_{\alg}^{(k)},\wh\cG_{G,{\alg}}^{(k)}) -
\mL(\bZ^{(k)};\wh{\bbeta}^{(k)},\wh\cG_{G}^{(k)}) = o_p(1/T)$ for $G\geq G_0$.
Under the conditions of Theorem~\ref{thm:consistency2}, the lower bound \eqref{lower_bound_Q} remains valid.
\end{corollary}

\begin{remark}
The above theoretical analysis explains the rationale of our data splitting mechanism on the time span in Figure \ref{fig:split}. 
Particularly, since the  two data folds share the same group structure $\cG_G$, we can successfully link the dominating term \eqref{eq:dom} to the optimization on the training data fold $\Z^{(1)}$. 
We further connect this optimization to the maximization of \eqref{eq:Delta_g1g2} with respect to $\{\mC_g^1, \mC_g^2\}$ to establish the lower bound. 
This is the key ingredient for analyzing the overfitting case. 
On the other hand, if we split the panel data on the individuals (denoted as $\mZ^{(1)} = \{\z_{it}: i\in [\lfloor N/2 \rfloor], t\in [T]\}$ and $\mZ^{(2)} = \Z\backslash \mZ^{(1)}$), we need to re-estimate the memberships on $\mZ^{(2)}$ with $\wh \bbeta^{(1)}$ obtained from $\mZ^{(1)}$. 
Therefore, it does not involve an optimization procedure to seek the best partition for the individuals on the testing data fold.
    A similar phenomenon is also observed for estimating the clustering number of clustering tasks, which is known as the ``double dipping'' problem in the clustering literature \citep{chen2023selective,yun2023selective,gao2024selective}. 
    We also compare our theoretical frameworks with the CV method used in change point detection of \cite{zou2020consistent}, and we refer to Remark \ref{rem:comp_zou} in  Appendix \ref{subsection:proof_of_thm_consistency1} for details.
    \label{rem:split}
\end{remark}

\vspace{-0.5cm}
\section{Group Panel Data Model with Fixed Effects}\label{sec:model_fixed}

\subsection{Panel Data Estimation with Fixed Effects}\label{subsection:panel_data_est_with_alpha}

In this section, we further discuss group number estimation based on the cross-validation method for nonlinear panel data models with fixed effects. 
In practice, individual-level heterogeneity may exist,  and the cross-validation method needs to be revised to accommodate the individual-specific heterogeneity.
Suppose the fixed effect of individual $i$ is denoted by $\alpha_i$, which characterizes
individual-level heterogeneity. 
With a slight abuse of notation, throughout this section, we use
$\boldsymbol{\ell}(\bz_{it};\bbeta_{g_i},\alpha_i)$ and
$\mL(\bZ;\bbeta,\balpha,\cG_G)$
to denote the per-observation loss and the total loss when the individual
fixed effects $\{\alpha_i\}$ are present.
We can obtain parameter estimation
by minimizing the loss function $\mL(\bZ;\bbeta,\balpha,\cG_G)$ with a pre-specified group number $G$, i.e.,
\compressmath
\begin{align}
\big\{\wh \bbeta,\wh \balpha,\wh \cG_G \big\} = \argmin_{\bbeta,\balpha,\cG_G}\mL(\bZ;\bbeta,\balpha,\cG_G)=\argmin_{\bbeta,\balpha,\cG_G}\frac{1}{NT} \sum_{i =1}^N\sum_{t = 1}^T\boldsymbol{\ell}(\bz_{it}; \bbeta_{g_i},\alpha_i),\label{equ:est_with_fe}
\end{align}
where $\balpha = (\alpha_1,\alpha_2,\ldots, \alpha_N)^\top \in \mR^N$ and $\boldsymbol{\ell}(\z_{it};\bbeta_{g_i},\alpha_i)$ represents the loss function on $\z_{it}$ with parameters $\bbeta_{g_i}$ and $\alpha_i$.
Here we treat $\{\alpha_i\}$ as fixed effects in our analysis, which is also used in panel data literature \citep{bonhomme2015grouped,liu2020identification}.
%Here we abuse the notation slightly by using the same $\mL(\cdot)$.
In practice, we can use a profile objective function for estimation.
Specifically, define $\wh \alpha_i(\bbeta) = \argmin_{\alpha}
T^{-1}\sum_{t=1}^T\boldsymbol{\ell}(\bz_{it};\bbeta, \alpha)$ and then the profile objective function is given by
\begin{align}
\mL^\p(\bZ;\bbeta,\cG_G)=\frac{1}{N}\sum_{i = 1}^N\ol{\boldsymbol{\ell}}_i^\p(\bZ;\bbeta_{g_i})\defeq
\frac{1}{NT} \sum_{i =1}^N\sum_{t = 1}^T\boldsymbol{\ell}\big(\bz_{it}; \bbeta_{g_i}, \wh\alpha_i(\bbeta_{g_i})\big).\label{Lp_loss}
\end{align}
A feasible iterative algorithm for obtaining
$\{\wh \bbeta, \wh \cG_G \}= \argmin_{\bbeta, \cG_G}\mL^\p(\bZ;\bbeta,\cG_G)$ is summarized in Algorithm \ref{estimation_algo_fixed_effect}.
Subsequently, we discuss the estimation of $G_0$ in the presence of the fixed effects
$\{\alpha_i:1\le i\le N\}$ based on the profile objective function.
We define $\boldsymbol{\ell}^{\bbeta}(\bz_{it};\bbeta, \alpha_i) = \partial
\boldsymbol{\ell}(\bz_{it}; \bbeta, \alpha_i)/\partial \bbeta$, and
$\boldsymbol{\ell}^\alpha(\bz_{it}; \bbeta, \alpha_i) = \partial
\boldsymbol{\ell}(\bz_{it}; \bbeta, \alpha_i)/\partial \alpha_i$.
Similarly, define $\boldsymbol{\ell}^{\alpha{\bbeta}}(\bz_{it}; \bbeta, \alpha_i)$
and $\boldsymbol{\ell}^{\alpha\alpha}(\bz_{it}; \bbeta, \alpha_i)$ and let
$\ol \mL_i^{\alpha{\bbeta}}(\bbeta, \alpha_i) = T^{-1}\sum_{t=1}^T\boldsymbol{\ell}^{\alpha{\bbeta}}(\bz_{it}; \bbeta, \alpha_i) $ and
$\ol \mL_i^{\alpha\alpha}(\bbeta, \alpha_i) = T^{-1}\sum_{t=1}^T\boldsymbol{\ell}^{\alpha\alpha}(\bz_{it}; \bbeta, \alpha_i) $.
Following the literature \citep{liu2020identification}, denote
\compressmath
\begin{align}
U_i(\bz_{it}; \bbeta, \alpha_i) &= \boldsymbol{\ell}^{\bbeta}(\bz_{it}; \bbeta, \alpha_i) -
   \boldsymbol{\ell}^\alpha(\bz_{it}; \bbeta, \alpha_i)
   \ol \mL_i^{\alpha\alpha} ( \bbeta, \alpha_i)^{-1}
   \ol \mL_i^{\alpha{\bbeta}} (\bbeta, \alpha_i);\label{def_Ui}\\
U_i^{\bbeta}(\bz_{it}; \bbeta, \alpha_i) &= \frac{\partial U_i(\bz_{it}; \bbeta, \alpha_i)}{\partial \bbeta},\qquad U_i^{\bbeta\bbeta}(\bz_{it}; \bbeta, \alpha_i) = \frac{\partial^2 U_i(\bz_{it}; \bbeta, \alpha_i)}{\partial \bbeta^2};\nonumber\\
{\small U_i^{\alpha}(\bz_{it}; \bbeta, \alpha_i)} &= {\small \frac{\partial U_i(\bz_{it}; \bbeta, \alpha_i)}{\partial \alpha}, U_i^{\alpha\alpha}(\bz_{it}; \bbeta, \alpha_i) = \frac{\partial^2 U_i(\bz_{it}; \bbeta, \alpha_i)}{\partial \alpha^2};U_i^{\alpha\bbeta}(\bz_{it}; \bbeta, \alpha_i) = \frac{\partial^2 U_i(\bz_{it}; \bbeta, \alpha_i)}{\partial \alpha\partial\bbeta}}.\nonumber
\end{align}
Denote $U_i(\bz_{it})=U_i(\bz_{it}; \bbeta_{g_i^0}^0, \alpha_i^0)$.
In addition, define $U_i^{\bbeta}(\bz_{it}) = \partial U_i(\bz_{it}; \bbeta_{g_i^0}^0, \alpha_i^0)/\partial \bbeta$
and $U_i^{\alpha}(\bz_{it})$, $U_i^{\bbeta\bbeta}(\bz_{it})$, $U_i^{\alpha\bbeta}(\bz_{it})$, $U_i^{\alpha\alpha}(\bz_{it})$, $\mL_i(\bz_{it})$ in the same way at the true value. %and $\ol U_i(\bbeta,\alpha_i)=T^{-1}\sum_{t=1}^T U_i(\bz_{it};\bbeta,\alpha_i)$.
Specifically, we use $\ol U_i(\bbeta,\alpha_i)$ to denote $T^{-1}\sum_{t=1}^T U_i(\bz_{it};\bbeta,\alpha_i)$ and use $\wt U_i$ to denote $T^{-1}\sum_{t=1}^T U_i(\bz_{it};\bbeta_{g_i^0}^0,\alpha_i^0)$.
Similarly, define $\wt \mL_i^{\alpha}$,
$\wt \mL_i^{\alpha\alpha}$,
$\wt U_i^{\alpha\alpha}$,
$\wt U_i^{\bbeta}$,
$\wt U_i^{\alpha\bbeta}$,
$\wt U_i^{\bbeta\bbeta}$
in the same way at the true value.
Given the group memberships $\cG_G$, we can define $\boldsymbol{\ell}^\p(\bz_{it};\bbeta)=\boldsymbol{\ell}(\bz_{it};\bbeta,\wh\alpha_i(\bbeta))$ and verify that
\compressmath
\begin{align*}
\sum_{i\in \mC_g}\sum_{t =1}^T
\frac{\partial \boldsymbol{\ell}^\p(\bz_{it};\bbeta_{g})}{\partial \bbeta_g} =
\sum_{i\in \mC_g}\sum_{t =1}^TU_i\big(
\bz_{it}; \bbeta_g, \wh\alpha_i(\bbeta_g)\big).
\end{align*}
To derive a simple yet effective loss function when the fixed effects are present, we need the following result on estimation properties. The technical conditions are listed in Appendix \ref{subsection:technical_conditions_for_fixed_effects_model}.

%\begin{algorithm}[H]
%\footnotesize
%\setstretch{1.2}
%\caption{\footnotesize A $k$-means Type Algorithm for Group Panel Models with Fixed Effects}\label{estimation_algo_fixed_effect}
%\begin{algorithmic}[1]
%\Require
%The number of groups $G$;
%the dataset $\bZ$.
%\Ensure
%The estimation $\widehat{\bbeta}$ and $\wh\cG_G$.
%\State Choose the initial estimators $\wh\bbeta^{(0)}= (\wh\bbeta_1^{(0)},\cdots,\wh\bbeta_G^{(0)\top})\in \mR^{G \times p}$.
%\Repeat
%\State In step $s+1$:
%\State \textbf{Update group membership}: For $i=1,\ldots,N$,
%$\wh g_{i}^{(s+1)}=
%\arg\min_{g \in[G]} 1/T \sum_{t = 1}^T\mL^\p \big(\bz_{it}; \wh\bbeta_{\wh g_{i}^{(s)}}^{(s)}\big).$
%\State \textbf{Update coefficients}: $\wh \bbeta^{(s+1)} = \argmin_{\bbeta\in \mR^{G \times p} }\mL^\p (\bZ; \bbeta, \wh\cG_G^{(s+1)}).$
%\Until{the convergence criterion is met.}
%\State Return $\widehat{\bbeta}$ and $\wh\cG_G$.
%\end{algorithmic}
%\end{algorithm}

\begin{algorithm}[H]
\footnotesize
\setstretch{1.2}
\caption{A $k$-means Type Algorithm for Group Panel Models with Fixed Effects}
\label{estimation_algo_fixed_effect}
\begin{algorithmic}[1]
\Require Number of groups $G$; dataset $\bZ$; loss function $\ol{\boldsymbol{\ell}}_i^{\p}(\cdot;\cdot)$;
         tolerance $\epsilon$; maximum number of iterations $S_{\max}$.
\Ensure  Estimators $\wh\bbeta$ and $\wh\cG_G$.
\State Initialize $\wh\bbeta^{(0)}=(\wh\bbeta_1^{(0)},\ldots,\wh\bbeta_G^{(0)})^\top\in\mR^{G\times p}$; set $s\gets 0$.
\Repeat
   \State \underline{\textbf{Update group membership.}} For $i=1,\ldots,N$,\quad$\wh g_i^{(s+1)} = \argmin_{g\in[G]} \ol{\boldsymbol{\ell}}_i^{\p}\big(\bZ;\wh\bbeta_g^{(s)}\big).$
   \Statex \qquad Set $\wh\mC_g^{(s+1)}=\{i:\wh g_i^{(s+1)}=g\}$ for $g\in[G]$, and $\wh\cG_G^{(s+1)}=\{\wh\mC_1^{(s+1)},\ldots,\wh\mC_G^{(s+1)}\}$.
   \State \underline{\textbf{Update coefficients.}}  For $g=1,\ldots,G$, \quad $      \wh\bbeta_g^{(s+1)} = \argmin_{\bbeta_g\in\mR^{p}}\sum_{i\in\wh\mC_g^{(s+1)}}\ol{\boldsymbol{\ell}}_i^{\p}\big(\bZ;\bbeta_g\big).$
   \State $s\gets s+1$.
\Until{$\wh g_i^{(s)}=\wh g_i^{(s-1)}$ for all $i$, or $\|\wh\bbeta^{(s)}-\wh\bbeta^{(s-1)}\|_F<\epsilon$, or $s\ge S_{\max}$.}
\State \Return $\wh\bbeta=\wh\bbeta^{(s)}$ and $\wh\cG_G=\wh\cG_G^{(s)}$.
\end{algorithmic}
\end{algorithm}

\begin{algorithm}[H]
\footnotesize
\setstretch{1.2}
\caption{\footnotesize A Two-step Algorithm for Group Panel Models with Fixed Effects}\label{estimation_algo_fixed_effect_twostep}
\begin{algorithmic}[1]
\Require
The number of groups $G$;
the dataset $\bZ$.
\Ensure
Estimators $\widehat{\bbeta}$ and $\wh\cG_G$.
\State Use Algorithm \ref{estimation_algo_fixed_effect} to obtain estimators $\wh\bbeta^{\star}$ and $\wh\cG_G^{\star}$.
\State Use $\{\wh\bbeta^{\star},\wh\cG_G^{\star}\}$ to construct $\wh \bV_i$ by \eqref{equ:weight_cal}.
\State Solve $\{\wh\bbeta,\wh\cG_G\}=\argmin_{\bbeta,\cG_G}\mQ(\bZ;\bbeta,\cG_G)$, where $\mQ(\bZ;\bbeta,\cG_G)$ is defined in \eqref{Q_loss_fe}.
\State Return $\widehat{\bbeta}$ and $\wh\cG_G$.
\end{algorithmic}
\end{algorithm}

\begin{algorithm}[H]
\footnotesize
\setstretch{1.2}
\caption{\footnotesize Group Number Selection for Group Panel Models with Fixed Effects}\label{selection_algo_fixed_effect}
\begin{algorithmic}[1]
\Require
The maximum possible number of groups $G_{\max}$;
the dataset $\bZ$;
buffer zone length $\tau_{NT}$.
\Ensure
Selected number of groups $\wh G$.
\State Split $\bZ$ into $\bZ^{(1)}$ and $\bZ^{(2)}$ with a buffer zone of length $\tau_{NT}$ as in Figure~\ref{fig:split}.
\For{$G=1,2,\ldots, G_{\max}$}
\State Use Algorithm \ref{estimation_algo_fixed_effect_twostep} separately on $\bZ^{(1)}$ and $\bZ^{(2)}$ to obtain $\{\wh\bbeta^{(1)},\wh\cG_G^{(1)}\}$ and $\{\wh\bbeta^{(2)},\wh\cG_G^{(2)}\}$.
\State Calculate the validation loss $\ol \mQ(G)\defeq \mQ(\bZ^{(2)}; \wh{\bbeta}^{(1)}, \wh \cG_G^{(1)})+\mQ(\bZ^{(1)}; \wh{\bbeta}^{(2)}, \wh \cG_G^{(2)}).$
\EndFor
\State \Return $\wh G = \argmin_{G\in[G_{\max}]} \ol\mQ(G)$.
\end{algorithmic}
\end{algorithm}

\bep\label{prop:Ui_expan}
Assume conditions \eqref{condition:param_space_fixed}--\eqref{condition:identification_fixed} in Appendix \ref{subsection:technical_conditions_for_fixed_effects_model} and \eqref{condition:distribution},   \eqref{condition:sample_size}--\eqref{condition:group_separate} hold. 
In addition, define
$\bV_i = E\{\partial U_i(\bz_{it}; \bbeta_g^0, \alpha_i^0)/
\partial \bbeta^\top\}$.
Then we have
\compressmath
\begin{align}
\ol U_i(\bbeta_g, \wh\alpha_i(\bbeta_g)) =  \bV_i(\bbeta_g - \bbeta_g^0)+\wt U_i
+R_i+o_p(T^{-1}) + o_p(\|\bbeta_g - \bbeta_g^0\|_2),\label{U_i_avg_expan_prop}
\end{align}
for any $\bbeta_g$ satisfying $\|\bbeta_g-\bbeta_g^0\|_2=o_p(1)$, where
\compressmath
\begin{align}
R_i \defeq \Big[\frac{\wt\mL_i^{\alpha}}{E(\wt\mL_i^{\alpha\alpha})}\Big]
\Big[\frac{E(\wt U_i^{\alpha\alpha})\wt\mL_i^{\alpha}}
{2E(\wt \mL_i^{\alpha\alpha})}-
\wt U_i^\alpha\Big].\label{R_expr}
\end{align}
\eep

\noindent
The proof of Proposition \ref{prop:Ui_expan} is given in Appendix \ref{subsection:proof_of_prop1_and_prop2}.
First, the leading term involves both $\wt U_i$ and $R_i$.
Particularly, the $R_i$ term is an extra bias term caused by the individual-level fixed effects, which is in the order of $O_p(T^{-1})$.
It cannot be reduced by using aggregated information of all individuals.
The bias term will disappear when the fixed effects are not present.
Second, as suggested by the expansion of $\ol U_i(\bbeta_g, \wh\alpha_i(\bbeta_g))$, an individual weighting matrix $\bV_i$ is involved in the linear leading term $\bbeta_g - \bbeta_g^0$.
The weighting matrix varies across $i$ due to the existence of the
fixed effects.
Motivated by this fact,
we consider a revised weighted quadratic objective function
as
\compressmath
\begin{align}
\mQ(\bZ;\bbeta,\cG_G) = \frac{1}{N}\sum_{g = 1}^G
\sum_{i \in \mC_g} \ol U_i\big(
\bbeta_g, \wh\alpha_i(\bbeta_g)\big)^\top {\wh\bV_i^{-2}}%\bV_i^{-1}
\ol U_i\big(\bbeta_g, \wh\alpha_i(\bbeta_g)\big).\label{Q_loss_fe}
\end{align}
The $\mQ$-function in \eqref{Q_loss_fe} is actually a reweighted loss function after adjusting to the individual-level heterogeneous weighting matrix $\wh\bV_i$. 
Based on this $\mQ$-function, we obtain the final estimators by $\{\wh \bbeta,\wh\cG_G\} = \argmin_{\bbeta,\cG_G}\mQ(\bZ;\bbeta,\cG_G)$.

\subsection{Group Number Estimation using Cross-Validation Method}

Based on the objective function $\mQ(\bZ;\bbeta,\cG_G)$, we apply the same data-splitting scheme to estimate the number of groups.
Specifically, the estimated $\wh G$ is given by
\compressmath
\begin{align}
\wh G = {\argmin_{G\in[G_{\max}]}}\Big\{
\mQ(\bZ^{(2)}; \wh{\bbeta}^{(1)}, \wh \cG_G^{(1)}) +
\mQ(\bZ^{(1)}; \wh{\bbeta}^{(2)}, \wh \cG_G^{(2)})\Big\},\label{G_est_fe}
\end{align}
where $\{\wh\bbeta^{(1)}, \wh \cG_G^{(1)}\}$,
$\{\wh\bbeta^{(2)}, \wh \cG_G^{(2)}\}$ are estimators obtained by applying Algorithm~\ref{estimation_algo_fixed_effect_twostep} on two data folds respectively.
The  selection procedure is summarized in Algorithm \ref{selection_algo_fixed_effect}.
In what follows, we establish the estimation consistency of $\wh G$ under a pivotal condition (i.e., \eqref{lower_bound_Q_fixed_effects}) in Theorem \ref{thm:consistency_fixed_effects} and verify the condition in Theorem \ref{thm:consistency_fixed_effects2}.

\bet\label{thm:consistency_fixed_effects}
Suppose conditions \eqref{condition:param_space_fixed}-\eqref{condition:sample_size_fixed}  in Appendix  \ref{subsection:technical_conditions_for_fixed_effects_model} and \eqref{condition:distribution}, \eqref{condition:buffer_len}, \eqref{condition:group_size}, \eqref{condition:group_separate} hold.
Further assume for $G>G_0$,
\begin{align}
\mQ(\bZ^{(k)};\wh{\bbeta}^{(k)},\cG_{G_0}^0) -
	\mQ(\bZ^{(k)};\wh{\bbeta}^{(k)},\wh\cG_{G}^{(k)}) \gtrsim T^{-1}
\label{lower_bound_Q_fixed_effects},
\end{align}
for $k = 1,2$ with probability  tending to one.
Then we have $\lim_{\min(N,T)\rightarrow \infty} P(\wh G = G_0) = 1$.
\eet

The proof of Theorem \ref{thm:consistency_fixed_effects} is provided in Appendix \ref{subsection:proof_of_thm_consistency_fixed_effects}.
The condition \eqref{lower_bound_Q_fixed_effects} means that
the reduction in the $\mQ$-function due to further partitioning
$\cG_{G_0}^0$ by $\wh{\cG}_G^{(k)}$ (with $G>G_0$)
should also be at least equal to the rate $T^{-1}$, which is the same as \eqref{lower_bound_Q}.
In the following, we show that the lower bound condition \eqref{lower_bound_Q_fixed_effects} can be satisfied by using the revised $\mQ$-function.

\bet\label{thm:consistency_fixed_effects2}
Suppose conditions \eqref{condition:param_space_fixed}-\eqref{condition:sample_size_fixed}  in Appendix  \ref{subsection:technical_conditions_for_fixed_effects_model} and \eqref{condition:distribution}, \eqref{condition:buffer_len}, \eqref{condition:group_size}, \eqref{condition:group_separate} hold.
Then the condition \eqref{lower_bound_Q_fixed_effects} is valid.
\eet

The proof of Theorem \ref{thm:consistency_fixed_effects2} is provided in Appendix \ref{subsection:proof_of_thm_consistency_fixed_effects2}.
The conclusion of Theorem \ref{thm:consistency_fixed_effects2} ensures that \eqref{lower_bound_Q_fixed_effects} holds automatically.
As an immediate consequence of Theorem \ref{thm:consistency_fixed_effects}, we can conclude that $\lim_{\min(N,T)\rightarrow \infty} P(\wh G = G_0) = 1$ still holds for the group panel data model with fixed effects.
Following the similar routine of the overfitting part in Theorem \ref{thm:consistency2}, we show that the lower bound in
\eqref{lower_bound_Q_fixed_effects} can be guaranteed by the optimization procedure.
We also refer to Appendix \ref{sec:model_interactive} for two possible extensions of our proposed method.

\section{Numerical Studies}\label{sec:numerical_study}

\subsection{Simulation Models and Selection Criteria}\label{subsec:simulation_model}

To evaluate the finite sample performance of the CV method, we conduct a number of simulation studies in
this section.
{For comparison, we include several competing methods discussed in Section  \ref{sec:comp_IC}.
We refer to the information criteria given by \cite{bonhomme2015grouped}, \cite{su2016identifying}
and \cite{liu2020identification} as BIC, LIC (Lasso-based information criterion) and  PC (penalty criterion), respectively in the following.
In addition to the IC based methods, we also consider the method based on hypothesis testing (denoted as HT)  proposed by \cite{lu2017determining}.}
%Note that the HT method is only designed for linear models.
%Therefore, we only include HT method for linear panel models (i.e., DGP 1 and DGP 2) in the following examples.
Since the HT method is designed only for linear models, we include it only for DGP 1 and DGP 2.

We consider four data generating processes (DGPs), including linear static (DGP 1) and dynamic panel models (DGP 2), a dynamic probit panel model (DGP 3), and a static Poisson panel model (DGP 4).
%The sample sizes are given as $N=80,120$ and $T=80,120,160$.
%Next, we set $G_0 = 4$ with equal group sizes as $N/G_0$.
The sample sizes are set to $N \in\{80,120\}$ and $T \in\{80,120,160\}$. 
The true number of groups is $G_0=4$, with equal group sizes $N / G_0$.
The detailed specifications of the four DGPs (i.e., DGP 1--DGP 4) are deferred to Appendix \ref{subsec:res_dgp1_4} due to space limitations.

\subsection{Simulation Results}

The random experiments are repeated for $R = 500$ times.
To measure the finite sample performance, we report the accuracy as Acc$=R^{-1}\sum_{r=1}^R\mathbb{I}(\wh G^{(r)}=G_0)$, the mean estimation bias as Bias$ = R^{-1}\sum_{r=1}^R(\wh G^{(r)}-G_0)$
and the root mean squared error as RMSE$ = \{R^{-1}\sum_{r=1}^R(\wh G^{(r)}-G_0)^2\}^{1/2}$, where $\wh G^{(r)}$ denotes the estimated group number in the $r$th replicate.
The numerical results for all four DGPs are summarized in Table \ref{simulation_res}.
For visualization, the distribution of $\wh G - G_0$ under DGP 2 (dynamic linear panel model with fixed effect $\alpha_i$), DGP 3 (dynamic probit panel model), and DGP 4 (static Poisson panel model) is presented in Figure \ref{fig:dgp_3_and_4}, while the corresponding plots for DGP 1 (static linear panel model with and without fixed effects) and DGP 2 (dynamic linear panel model without fixed effects) are deferred to Appendix \ref{subsec:res_dgp1_4} due to space limitations.

First, the estimation accuracy of the CV method improves substantially as $N$ and $T$ increase.
To further illustrate the consistency property established in Theorems \ref{thm:consistency1} and \ref{thm:consistency_fixed_effects}, we report  the selection accuracy of CV over a wider range of $(N,T)$ combinations in Figure \ref{fig:consistency_plot} in Appendix \ref{subsec:res_dgp1_4}, where the accuracy approaches $1$ as $N$ and $T$ grow large.
Besides, the proposed CV method performs better than, or is at least comparable to, the competing methods across the considered DGPs. 
The BIC, PC, and LIC methods severely underestimate $G_0$ with accuracies essentially equal to zero.
The HT method becomes comparable to CV only for linear panel models with sufficiently large $T$, but is inferior to CV when $T$ is small or moderate, and
is not applicable to the nonlinear models as in DGP 3 and DGP 4.
Particularly, the advantage of the proposed CV method is more pronounced for the Poisson panel
model (in DGP 4). This implies that
the IC methods are sensitive to tuning parameter specifications and requires appropriate tuning to deliver a good performance. In contrast, our
CV method does not involve this tuning process and can deliver a more robust performance with various
panel models.

\begin{figure}[!htbp]
\centering
\includegraphics[scale=0.48]{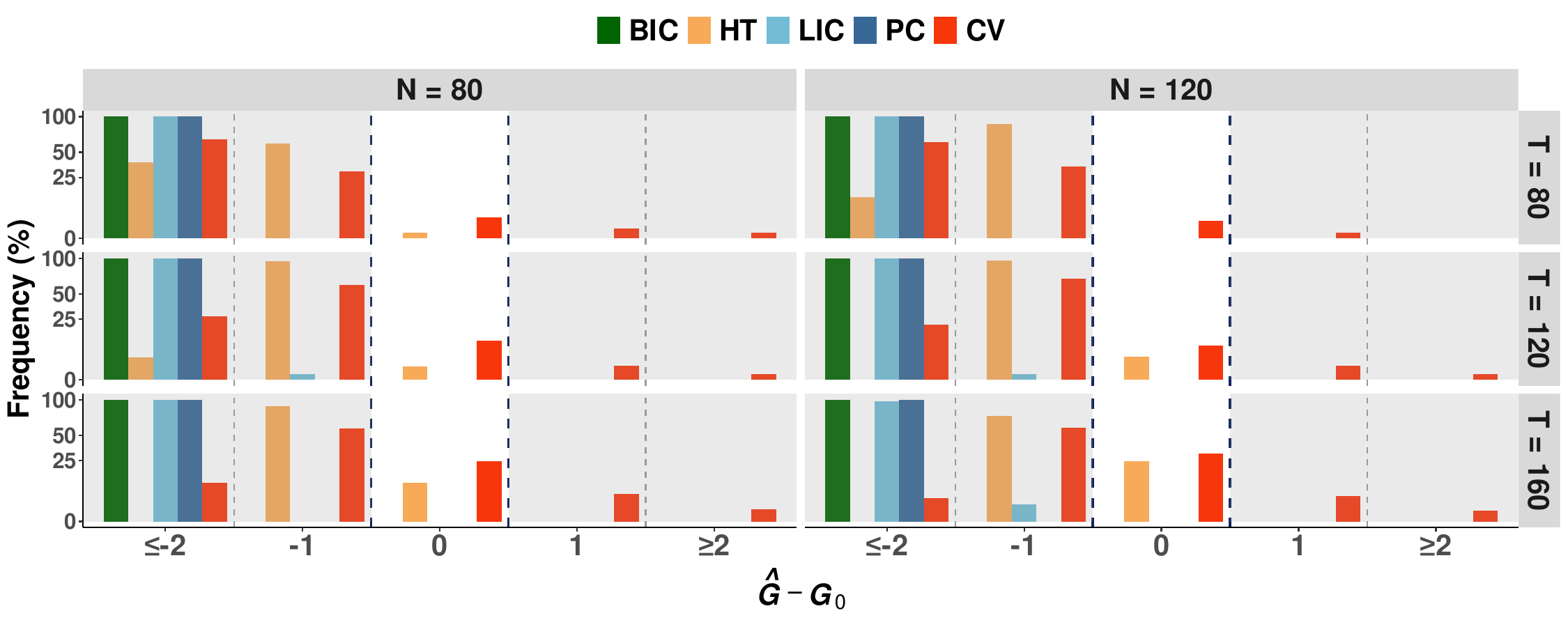}
  \includegraphics[scale=0.48]{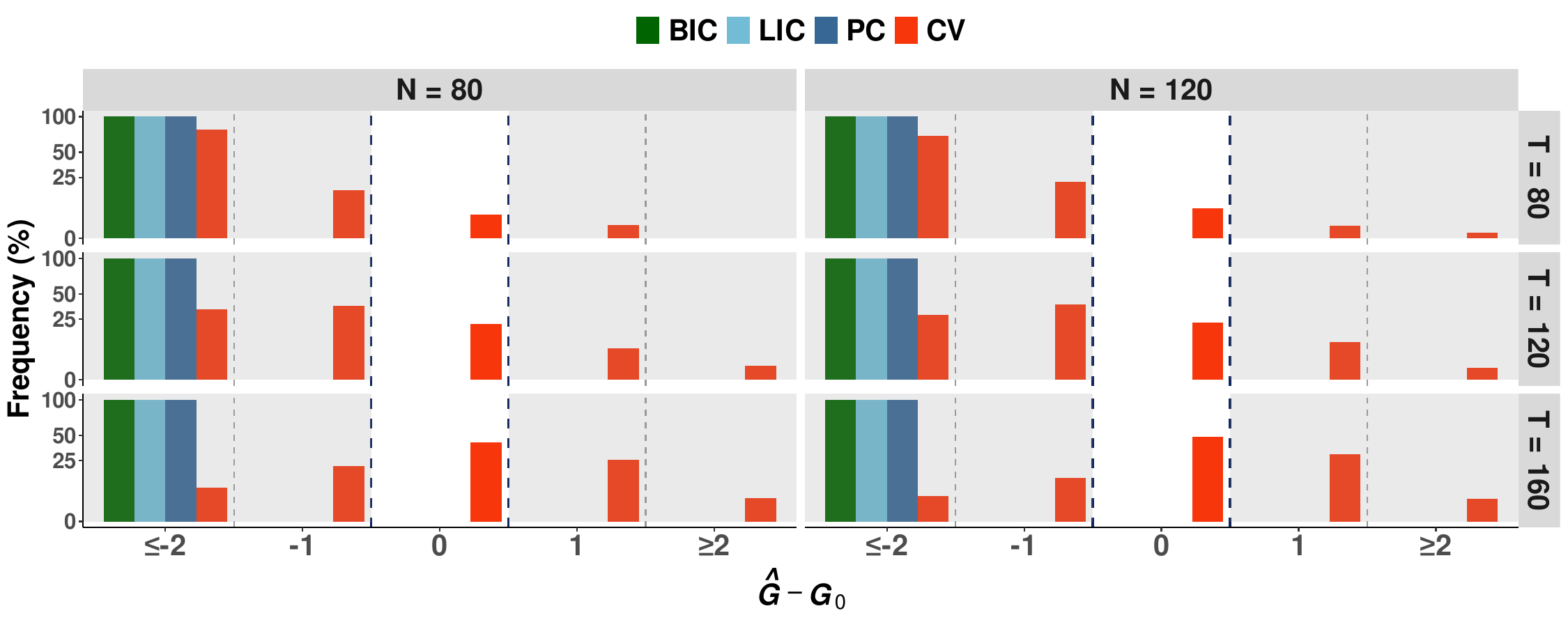}
  \includegraphics[scale=0.48]{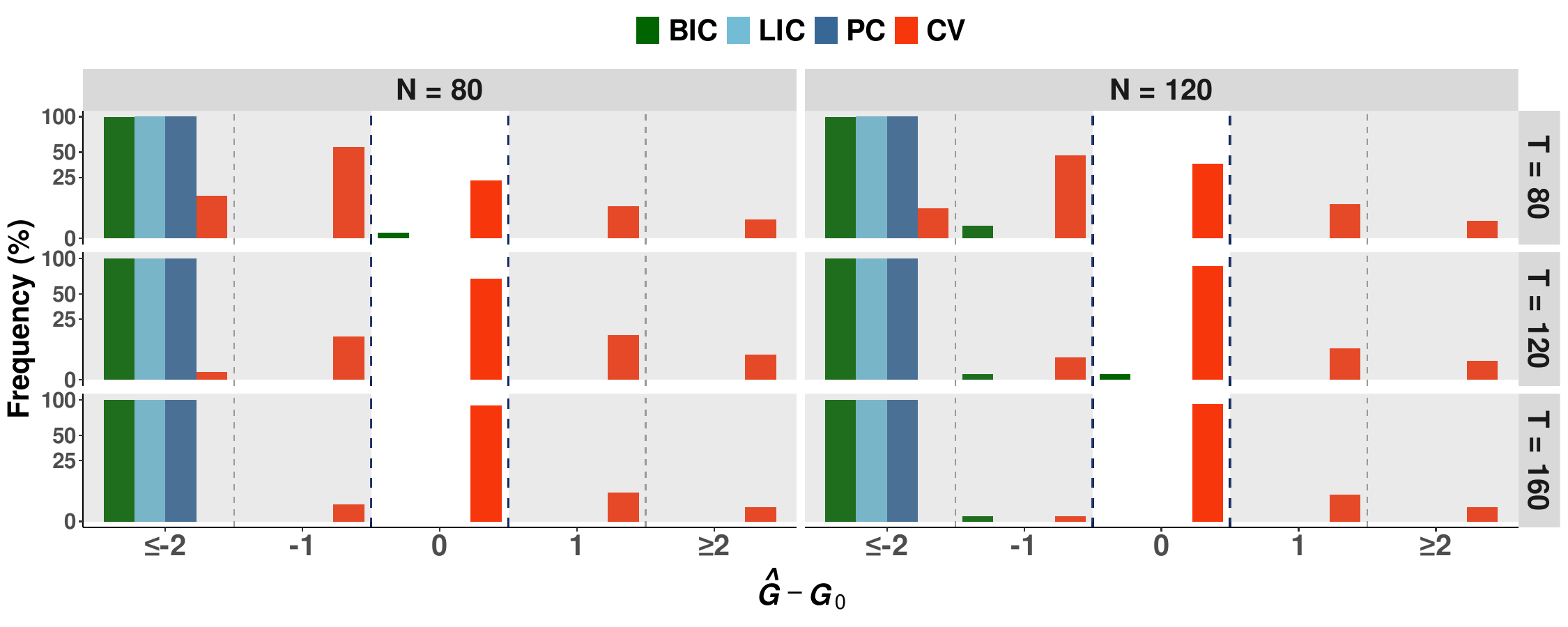}
  \caption{\small\it Distribution of $\wh G - G_0$ for the HT, PC, LIC, BIC and CV methods under DGP 2 (with fixed effect $\alpha_i$), DGP 3 and DGP 4.
  The top panel shows the performance under DGP 2 (with fixed effect $\alpha_i$), the middle panel shows the performance under DGP 3 and the bottom panel shows the performance under DGP 4.}\label{fig:dgp_3_and_4}
\end{figure}

\begin{landscape}
\newgeometry{left=2cm, right=-3cm, top=4cm, bottom=-1cm}
\begin{table}[thp]\footnotesize
\centering
\begin{minipage}[t]{8in}
\caption{Simulation results for all DGPs with 500 replications. The numerical performance is evaluated for different sample sizes $N$ and $T$. The Acc, Bias and RMSE are reported for all estimators under different DGPs.}\label{simulation_res}
\end{minipage}
\begin{tabular}{cccccccccccccccccccccc}
   \toprule
    \multirow{2}{*}{$N$} & \multirow{2}{*}{$T$} & \multirow{2}{*}{\textbf{Method}} & \multicolumn{3}{c}{DGP 1.} & \multicolumn{3}{c}{DGP 1. (with $\alpha_i$)} & \multicolumn{3}{c}{DGP 2.} & \multicolumn{3}{c}{DGP 2. (with $\alpha_i$)} & \multicolumn{3}{c}{DGP 3} & \multicolumn{3}{c}{DGP 4} \\
    \cmidrule(lr){4-6}\cmidrule(lr){7-9}\cmidrule(lr){10-12}\cmidrule(lr){13-15}\cmidrule(lr){16-18}\cmidrule(lr){19-21}
           &      &        & Acc   & Bias  & RMSE  & Acc   & Bias  & RMSE  & Acc   & Bias  & RMSE  & Acc   & Bias  & RMSE  & Acc   & Bias  & RMSE  & Acc   & Bias  & RMSE \\
    \midrule
    \multirow{15}{*}{80}
      & \multirow{5}{*}{80}
          & BIC & 0.00 & -2.95 & 2.95  & 0.00 & -2.99 & 2.99 & 0.00 & -3.00 & 3.00 & 0.00 & -3.00 & 3.00 & 0.00 & -2.05 & 2.06 & 0.00 & -2.00 & 2.00 \\
      &     & PC  & 0.00 & -3.00 & 3.00  & 0.00 & -3.00 & 3.00 & 0.00 & -3.00 & 3.00 & 0.00 & -3.00 & 3.00 & 0.00 & -3.00 & 3.00 & 0.00 & -3.00 & 3.00 \\
      &     & LIC & 0.00 & -1.62 & 1.69  & 0.00 & -1.75 & 1.80 & 0.00 & -2.00 & 2.00 & 0.00 & -2.00 & 2.00 & 0.00 & -3.00 & 3.00 & 0.00 & -3.00 & 3.00 \\
      &     & HT  & 0.00 & -1.03 & 1.07  & 0.00 & -1.03 & 1.07 & 0.00 & -1.39 & 1.47 & 0.00 & -1.39 & 1.47 & - & - & - & - & - & - \\
      &     & CV  & \textbf{0.07} & -1.23 & 1.42  & \textbf{0.08} & -1.16 & 1.41 & \textbf{0.02} & -1.66 & 1.75 & \textbf{0.03} & -1.61 & 1.72 & \textbf{0.02} & -2.27 & 2.43 & \textbf{0.23} & -0.68 & 1.11 \\
    \cmidrule(lr){2-3}\cmidrule(lr){4-6}\cmidrule(lr){7-9}\cmidrule(lr){10-12}\cmidrule(lr){13-15}\cmidrule(lr){16-18}\cmidrule(lr){19-21}
      & \multirow{5}{*}{120}
          & BIC & 0.00 & -2.98 & 2.99  & 0.00 & -3.00 & 3.00 & 0.00 & -3.00 & 3.00 & 0.00 & -3.00 & 3.00 & 0.00 & -2.06 & 2.07 & 0.00 & -2.00 & 2.00 \\
      &     & PC  & 0.00 & -3.00 & 3.00  & 0.00 & -3.00 & 3.00 & 0.00 & -3.00 & 3.00 & 0.00 & -3.00 & 3.00 & 0.00 & -3.00 & 3.00 & 0.00 & -3.00 & 3.00 \\
      &     & LIC & 0.00 & -1.37 & 1.45  & 0.00 & -1.48 & 1.56 & 0.00 & -2.00 & 2.00 & 0.00 & -2.00 & 2.00 & 0.00 & -3.00 & 3.00 & 0.00 & -3.00 & 3.00 \\
      &     & HT  & 0.12 & -0.87 & 0.95  & 0.11 & -0.88 & 0.95 & 0.02 & -1.02 & 1.04 & 0.01 & -1.02 & 1.04 & - & - & - & - & - & - \\
      &     & CV  & \textbf{0.26} & -0.65 & 1.09  & \textbf{0.28} & -0.59 & 1.04 & \textbf{0.10} & -1.13 & 1.32 & \textbf{0.10} & -1.14 & 1.31 & \textbf{0.10} & -1.56 & 1.78 & \textbf{0.69} & 0.10 & 0.71 \\
    \cmidrule(lr){2-3}\cmidrule(lr){4-6}\cmidrule(lr){7-9}\cmidrule(lr){10-12}\cmidrule(lr){13-15}\cmidrule(lr){16-18}\cmidrule(lr){19-21}
      & \multirow{5}{*}{160}
          & BIC & 0.00 & -3.00 & 3.00  & 0.00 & -3.00 & 3.00 & 0.00 & -3.00 & 3.00 & 0.00 & -3.00 & 3.00 & 0.00 & -2.08 & 2.09 & 0.00 & -2.00 & 2.00 \\
      &     & PC  & 0.00 & -3.00 & 3.00  & 0.00 & -3.00 & 3.00 & 0.00 & -3.00 & 3.00 & 0.00 & -3.00 & 3.00 & 0.00 & -3.00 & 3.00 & 0.00 & -3.00 & 3.00 \\
      &     & LIC & 0.00 & -1.17 & 1.23  & 0.00 & -1.24 & 1.31 & 0.00 & -2.00 & 2.00 & 0.00 & -2.00 & 2.00 & 0.00 & -3.00 & 3.00 & 0.00 & -3.00 & 3.00 \\
      &     & HT  & 0.44 & -0.55 & 0.77  & 0.42 & -0.56 & 0.78 & 0.11 & -0.89 & 0.94 & 0.10 & -0.90 & 0.95 & - & - & - & - & - & - \\
      &     & CV  & \textbf{0.59} & -0.20 & 0.76  & \textbf{0.57} & -0.13 & 0.81 & \textbf{0.22} & -0.74 & 1.07 & \textbf{0.25} & -0.72 & 1.05 & \textbf{0.34} & -0.82 & 1.23 & \textbf{0.91} & 0.07 & 0.38 \\
    \midrule
    \multirow{15}{*}{120}
      & \multirow{5}{*}{80}
          & BIC & 0.00 & -2.06 & 2.08  & 0.00 & -2.22 & 2.26 & 0.00 & -2.94 & 2.95 & 0.00 & -2.98 & 2.99 & 0.00 & -2.01 & 2.01 & 0.00 & -1.99 & 1.99 \\
      &     & PC  & 0.00 & -3.00 & 3.00  & 0.00 & -3.00 & 3.00 & 0.00 & -3.00 & 3.00 & 0.00 & -3.00 & 3.00 & 0.00 & -3.00 & 3.00 & 0.00 & -3.00 & 3.00 \\
      &     & LIC & 0.00 & -1.20 & 1.27  & 0.00 & -1.36 & 1.45 & 0.00 & -2.00 & 2.00 & 0.00 & -2.00 & 2.00 & 0.00 & -3.00 & 3.00 & 0.00 & -3.00 & 3.00 \\
      &     & HT  & 0.02 & -0.98 & 1.00  & 0.01 & -0.99 & 1.00 & 0.00 & -1.10 & 1.14 & 0.00 & -1.11 & 1.16 & - & - & - & - & - & - \\
      &     & CV  & \textbf{0.06} & -1.19 & 1.34  & \textbf{0.06} & -1.17 & 1.34 & \textbf{0.02} & -1.65 & 1.74 & \textbf{0.02} & -1.60 & 1.69 & \textbf{0.06} & -2.27 & 2.48 & \textbf{0.38} & -0.46 & 0.93 \\
    \cmidrule(lr){2-3}\cmidrule(lr){4-6}\cmidrule(lr){7-9}\cmidrule(lr){10-12}\cmidrule(lr){13-15}\cmidrule(lr){16-18}\cmidrule(lr){19-21}
      & \multirow{5}{*}{120}
          & BIC & 0.00 & -2.17 & 2.21  & 0.00 & -2.36 & 2.40 & 0.00 & -2.99 & 2.99 & 0.00 & -3.00 & 3.00 & 0.00 & -2.00 & 2.00 & 0.00 & -1.99 & 2.00 \\
      &     & PC  & 0.00 & -3.00 & 3.00  & 0.00 & -3.00 & 3.00 & 0.00 & -3.00 & 3.00 & 0.00 & -3.00 & 3.00 & 0.00 & -3.00 & 3.00 & 0.00 & -3.00 & 3.00 \\
      &     & LIC & 0.00 & -1.06 & 1.09  & 0.00 & -1.11 & 1.15 & 0.00 & -1.99 & 1.99 & 0.00 & -2.00 & 2.00 & 0.00 & -3.00 & 3.00 & 0.00 & -3.00 & 3.00 \\
      &     & HT  & 0.24 & -0.76 & 0.87  & 0.21 & -0.78 & 0.90 & 0.05 & -0.95 & 0.98 & 0.04 & -0.96 & 0.98 & - & - & - & - & - & - \\
      &     & CV  & \textbf{0.41} & -0.53 & 0.90  & \textbf{0.42} & -0.50 & 0.88 & \textbf{0.08} & -1.10 & 1.26 & \textbf{0.08} & -1.09 & 1.25 & \textbf{0.15} & -1.43 & 1.71 & \textbf{0.87} & 0.09 & 0.48 \\
    \cmidrule(lr){2-3}\cmidrule(lr){4-6}\cmidrule(lr){7-9}\cmidrule(lr){10-12}\cmidrule(lr){13-15}\cmidrule(lr){16-18}\cmidrule(lr){19-21}
      & \multirow{5}{*}{160}
          & BIC & 0.00 & -2.25 & 2.30  & 0.00 & -2.40 & 2.45 & 0.00 & -3.00 & 3.00 & 0.00 & -3.00 & 3.00 & 0.00 & -2.00 & 2.00 & 0.00 & -2.00 & 2.00 \\
      &     & PC  & 0.00 & -3.00 & 3.00  & 0.00 & -3.00 & 3.00 & 0.00 & -3.00 & 3.00 & 0.00 & -3.00 & 3.00 & 0.00 & -3.00 & 3.00 & 0.00 & -3.00 & 3.00 \\
      &     & LIC & 0.00 & -1.00 & 1.01  & 0.00 & -1.01 & 1.01 & 0.00 & -1.97 & 1.98 & 0.00 & -1.98 & 1.98 & 0.00 & -3.00 & 3.00 & 0.00 & -3.00 & 3.00 \\
      &     & HT  & 0.73 & -0.21 & 0.62  & 0.71 & -0.20 & 0.68 & 0.27 & -0.73 & 0.86 & 0.24 & -0.76 & 0.87 & - & - & - & - & - & - \\
      &     & CV  & \textbf{0.73} & -0.11 & 0.56  & \textbf{0.73} & -0.07 & 0.57 & \textbf{0.30} & -0.68 & 0.95 & \textbf{0.31} & -0.61 & 0.91 & \textbf{0.47} & -0.59 & 1.00 & \textbf{0.93} & 0.08 & 0.37 \\
    \bottomrule
\end{tabular}
\end{table}
\restoregeometry
\end{landscape}

{

\section{Empirical Study}\label{empirical_study_main}

In this section, we apply the proposed method to the Chinese A-share market dataset collected from CSMAR (\href{https://data.csmar.com/}{https://data.csmar.com/}).
Our study aims to investigate the heterogeneous group patterns in the Chinese stock market
from July 1, 2006 to November 30, 2009 including the
financial crisis in 2008, providing insights into market structure dynamics and systemic risk assessment.
To demonstrate the robustness of the proposed method, we also apply the CV method to a variety of synthetic datasets with extensive simulation studies; see Appendix \ref{sec:numerical_study_appendix} for details.

\subsection{Data Description}\label{empirical_data_description}

The empirical analysis employs data from the Chinese A-share market extracted from the CSMAR database. Our sample consists of 80 firms from the finance, properties, and commercial sectors that were continuously listed on either the Shanghai Stock Exchange or Shenzhen Stock Exchange from July 1, 2006 to November 30, 2009.

The dependent variable is the daily volatility, measured by the daily absolute return for company $i$ on the $t$th trading day
$
Y_{it}=|\log P_{it}-\log P_{i(t-1)} |,
$
where $P_{it}$ is the adjusted closing price of company $i$ on the $t$th trading day.
This measure of volatility is widely used in the literature as it captures the magnitude of daily price movements while being robust to the direction of price changes \citep{ding1993long,corsi2009simple}.
For the covariates, we employ a set of market microstructure indicators, i.e.,  the previous trading day's price range (defined as the difference between daily high and low prices), logarithmic trading volume \citep{xu2006time}, logarithmic turnover ratio \citep{barinov2014turnover}, and logarithmic closing price.

To investigate the structural changes in market behavior during the 2008 financial crisis, we follow \cite{ando2017clustering}
to partition the sample period into five distinct phases as
\begin{itemize}
\setlength{\itemsep}{0pt}
\setlength{\parskip}{0pt}
\setlength{\parsep}{0pt}
\item Period 1 (Pre-crisis): July 1, 2006 to December 31, 2006
\item Period 2 (Crisis onset): July 1, 2007 to December 31, 2007
\item Period 3 (Crisis peak): February 1, 2008 to August 31, 2008
\item Period 4 (Crisis aftermath): October 1, 2008 to March 31, 2009
\item Period 5 (Recovery): May 1, 2009 to November 30, 2009.
\end{itemize}
{To better understand the intrinsic data patterns within each period, we visualize cross-sectional correlation matrices of $\{Y_{it}\}$ across 80 firms for each period in Figure \ref{fig:period_plot}.
It shows that the stocks exhibit stronger correlations during the crisis periods 3 and 4.}
\begin{figure}[h!]
  \centering
  \includegraphics[width=\textwidth]{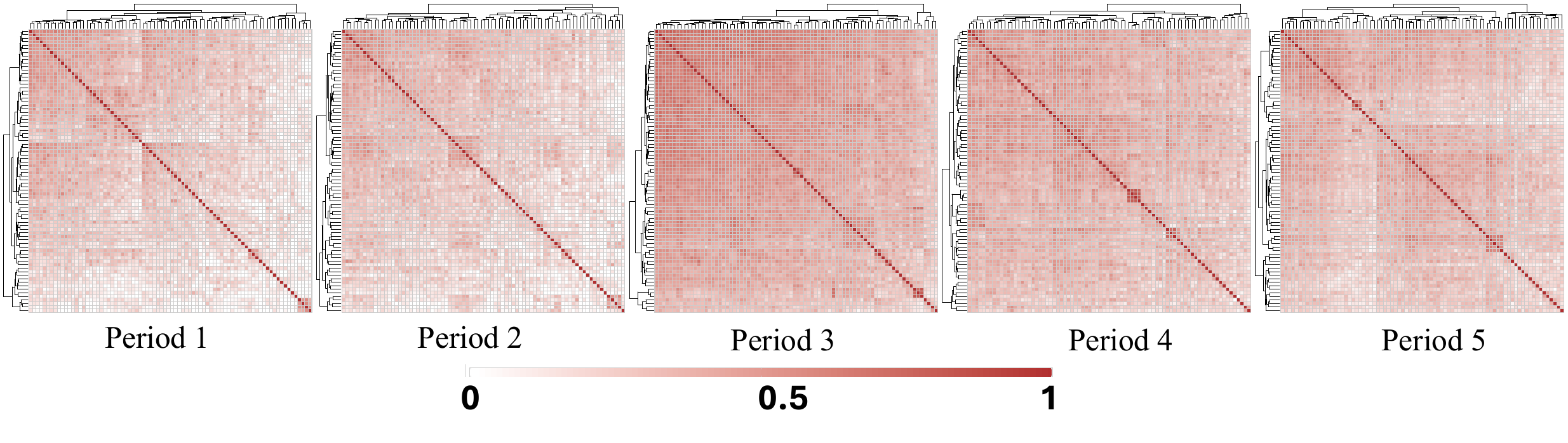}
  \caption{\small\it Correlation matrices of the response variable $\{Y_{it}\}$ across 80 firms during five periods.}\label{fig:period_plot}
\end{figure}
﻿

}
\subsection{Estimation Results}

Subsequently, we model the volatility using two linear panel models. First, we consider a linear panel model without fixed effects:
\begin{align}
Y_{it} = \gamma_{g_i} Y_{i(t-1)} + \bx_{it}^\top\bbeta_{g_i} + v_{it}, \label{equ:linear_noFE}
\end{align}
where \(v_{it}\) is an idiosyncratic error term. The autoregressive term \(\gamma_{g_i} Y_{i(t-1)}\) captures volatility persistence, while the group-specific slope \(\bbeta_{g_i}\) allows the sensitivities to market microstructure indicators.
Second, to account for unobserved time-invariant individual heterogeneity, we extend the model by incorporating individual fixed effects:
\begin{align}
Y_{it} = \alpha_i + \gamma_{g_i} Y_{i(t-1)} + \bx_{it}^\top\bbeta_{g_i} + v_{it}, \label{equ:linear_FE}
\end{align}
where \(\alpha_i\) denotes the firm-specific fixed effect reflecting firm-level characteristics such as industry attributes, ownership structure, and managerial style.

We then apply Algorithms \ref{selection_algorithm} and \ref{selection_algo_fixed_effect}  to estimate the number of groups within each period for Models \eqref{equ:linear_noFE} and \eqref{equ:linear_FE}, respectively. The estimation results are reported in Table \ref{tab:results}.
%\vspace{-5mm}
\begin{table}[htbp]
\centering
\caption{\it\small Group number estimation results in each period under Models \eqref{equ:linear_noFE} and \eqref{equ:linear_FE}.}
\begin{tabular}{cccccc}
\toprule
 & Period 1 & Period 2 & Period 3 & Period 4 & Period 5 \\
\midrule
Model \eqref{equ:linear_noFE} & 3 & 4 & 6 & 8 & 3 \\
Model \eqref{equ:linear_FE} & 1 & 2 & 2 & 1 & 1 \\
\bottomrule
\end{tabular}
\label{tab:results}
\end{table}
%\vspace{-5mm}
The results in Table \ref{tab:results} reveal distinctive patterns in the Chinese A-share market structure across different phases of the 2008 financial crisis.
For Model \eqref{equ:linear_noFE} without fixed effects, the estimated group number rises from 3 in Period 1 (pre-crisis) to 6 in Period 3 (crisis peak) and peaks at 8 in Period 4 (crisis aftermath), and lastly reduces to 3 in Period 5 (recovery).
This inverted-U pattern indicates that latent heterogeneity among firms intensifies substantially during the crisis, which is consistent with the strong cross-sectional correlations observed in Figure \ref{fig:period_plot} during Periods 3--4 and aligns with the findings of \cite{ando2017clustering} based on 31 global financial markets.
The reduction to 3 groups in Period 5 is consistent with a stabilization of market conditions after late-2008 policy interventions such as the four-trillion-yuan stimulus package launched in late 2008, reflecting the well-documented role of government intervention in shaping the stock market \citep{zhou2022government}.

Comparing the two models, Model \eqref{equ:linear_FE} consistently yields a substantially smaller number of groups than Model \eqref{equ:linear_noFE}, with at most 2 groups detected across all periods.
This contrast suggests that a large portion of the cross-sectional heterogeneity can be attributed to time-invariant firm-level characteristics.
Nevertheless, Model \eqref{equ:linear_FE} still detects more groups during the crisis (2 groups in Periods 2--3) than in the pre-crisis and recovery phases (1 group in Periods 1, 4, and 5), indicating that some residual crisis-related heterogeneity remains detectable around the onset and peak phases, although it is much weaker after controlling for firm fixed effects.

\section{Conclusion}\label{sec:conclusion}

To conclude the article, we provide several topics for future studies.
First, as suggested in Appendix \ref{sec:factor_model}, it is 
interesting to investigate the theoretical properties of the CV method 
under the interactive effects model. 
Second, it is worthwhile to extend the proposed method to the case of a 
diverging number of groups, and the corresponding theoretical properties 
can be developed accordingly.
Third, using the original loss $\mL$ directly as the selection 
criterion for the group number might also be workable, and we refer to 
Appendix \ref{appendix_original} for some numerical evidence. 
Theoretically, it remains to establish the selection consistency based 
on $\mL$, which would lead to a more convenient estimation procedure.
Lastly, a $K$-fold CV procedure can be employed to retain more training 
data and may exhibit improved finite-sample performance; it is 
challenging yet important to establish a valid CV procedure when 
$K\to\infty$.

\section*{Supplemental Material}

Appendices \ref{appendix_main_theoretical_properties} - \ref{sec:numerical_study_appendix}:
proofs of Theorem \ref{thm:consistency1} and Theorem \ref{thm:consistency2};
proofs of Theorem \ref{thm:consistency_fixed_effects} and Theorem \ref{thm:consistency_fixed_effects2};
{model extensions};
preliminary lemmas;
numerical studies.

\bibliographystyle{./asa}
\spacingset{1.14}
\bibliography{./reference}

\end{document}